\documentclass[11pt,a4paper]{article}
\usepackage{amsmath}
\usepackage[mathcal]{eucal}
\usepackage{latexsym}
\usepackage{amssymb}
\begin{titlepage}
\title{A systematic approach to the solution of the constraints of quantum gravity: The full theory.}
\author{Eyo Eyo Ita III}
\medskip
\input amssym.def
\input amssym.tex

\def \in{\indent}

\begin{document}
\maketitle
\bigskip
\centerline{Department of Applied Mathematics and Theoretical Physics} 
\smallskip
\centerline{Centre for Mathematical Sciences, University of Cambridge, Wilberforce Road}
\smallskip
\centerline{Cambridge CB3 0WA, United Kingdom}
\smallskip
\centerline{eei20@cam.ac.uk} 

\bigskip

\begin{abstract}
This is the third paper in a series outlining an algorithm to construct finite states of quantum gravity in Ashtekar variables.  In this paper we treat the case of the Klein--Gordon field quantized with gravity on the same footing.  We address the full theory, outlining the solution to the constraints and the construction of the corresponding wavefunction of the universe.  The basic method for the full theory is to expand the constraints relative to the solution for the pure Kodama state and rewrite them in the form of a generalized nonlinear group transformation of the CDJ matrix, viewed as a nine-dimensional vector.  We then outline a prescription for finding the fixed point of the flow, and the corresponding generalized Kodama state for the full theory is constructed.  The final solution is expressed in an asymptotic series in powers of model-specific matter inputs, suppressed by a small dimensionless constant, relative to the pure Kodama state.  We discuss this expansion from different perspectives.  Lastly, we explicitly show how the solution to the quantized constraints establishes a wavefunction of the universe with a predetermined semiclassical limit built in as a boundary condition on quantized gravity in the full theory.
\end{abstract}
\end{titlepage}

\section{Introduction}
A linearized treatment of general relativity in metric variables involves an expansion of the spacetime metric in fluctuations about a specified background metric.  For a flat Minkowski background, a ground state of metric relativity, the fluctuations have the interpretation of gravitons $h_{\mu\nu}$, given by

\begin{equation}
\label{MINK}
g_{\mu\nu}=\eta_{\mu\nu}+\sqrt{G}h_{\mu\nu}.
\end{equation}

\noindent
When (\ref{MINK}) is substituted into the action for general relativity, an infinite series is obtained in powers of the coupling constant $\sqrt{G}$.  It was first pointed out by Pauli that a loop expansion of the effective action of metric general relativity would be nonrenormalizable due to the negative mass dimension 
of $\sqrt{G}$.  In this paper we will try to make sense of an analogous expansion by treating the CDJ matrix $\Psi_{ae}$ as a metric in $SU(2)_{-}$ via the expansion

\begin{eqnarray}
\label{MINK1}
\Psi_{ae}=-{6 \over \Lambda}\bigl(\delta_{ae}+{\Lambda \over 6}\epsilon_{ae}\bigr).
\end{eqnarray}

\noindent
In (\ref{MINK1}) $\epsilon_{ae}$ is the CDJ deviation matrix, which parametrizes departure from the CDJ matrix for the pure Kodama state $\delta_{ae}$.  One difference between (\ref{MINK1}) and (\ref{MINK}) is the mass dimension of the coupling constants $([\Lambda]=2,[G]=-2)$.  We will ultimately suggest the hypothesis for a renormalizable description in Ashtekar variables due to the positive mass dimension of $\Lambda$.\par
\indent
It was argued in \cite{LINKO} that a linearization of the pure Kodama state about DeSitter spacetime is unnormalizable in the inner product of the theory of linearized gravitons in the Lorentzian theory.  However, it was also shown that the pure Kodama state $\Psi_{Kod}$ is the semiclassical manifestation of DeSitter spacetime \cite{KOD},\cite{POSLAMB}.  This begs the question, when one wishes to assess the suitability of the pure Kodama state $\Psi_{Kod}$ as a vacuum state of general relativity with a good semiclassical limit, as to whether one should rather expand fluctuations relative to $\Psi_{Kod}$ rather than expand $\Psi_{Kod}$ itself.\par
\indent
In \cite{WITTEN1} Witten argues that $\Psi_{Kod}$ cannot serve as the ground state for quantum gravity by analogy with Yang--Mills theory, the Chern--Simons state of which contains severe pathologies such as unnormalizability and nonunitarity.  However, Randono presents arguments in \cite{KODA1},\cite{KODA2} which addresses these issues when one interprets $\Psi_{Kod}$ essentially in analogy to plane wave momentum eigenstates.  Furthermore, Smolin and Soo in \cite{SOO} present the hypothesis for $\Psi_{Kod}$ as the ground state of quantum gravity with $\Lambda$ term by studying the behavior of small perturbations of $\Psi_{Kod}$ due to coupling of the theory to matter.  In this work the perturbations were measured relative to the self-dual sector for DeSitter spacetime, parametrized by $q^{ai}$

\begin{eqnarray}
\label{SELFDUAL}
\widetilde{q}^{ai}=\widetilde{E}^{ai}+{{3G} \over \lambda}\widetilde{B}^{ai}.
\end{eqnarray}

\noindent
It was shown for small perturbations that the matter degrees of freedom roughly correlate to $q^{ai}$ lending some credence to the notion of $\Psi_{Kod}$ as a proper ground state.  When one applies the (nondegenerate) Ashtekar magnetic field $B^i_a$ to (\ref{MINK1}) and uses the CDJ Ansatz \cite{EYO},\cite{THIE2}, one obtains

\begin{eqnarray}
\label{SELFDUAL1}
\widetilde{\sigma}^i_a+{6 \over \Lambda}B^i_a=-\epsilon_{ae}B^i_e.
\end{eqnarray}

\noindent
Equation (\ref{SELFDUAL1}) shows that the CDJ deviation 
matrix $\epsilon_{ae}$ is simply $q^{ai}$ with the spatial index $i$ projected into $SU(2)$.  As pointed out in \cite{SOO} it would be necessary to study the higher order corrections to the semiclassical limit in order to further investigate this hypothesis, but no method exists to do so.
\par
\indent
The ultimate purpose of this paper is to illustrate a new procedure by which the constraints of general relativity can be solved in the full theory, which enables the construction of the generalized Kodama state $\Psi_{GKod}$ \cite{EYO} devoid of ultraviolet singularities.  The claim is that $\Psi_{GKod}$ amounts to an expansion about $\Psi_{Kod}$ in terms of the matter content of the model.  Our new method will illustrate a systematic algorithm to obtain this expansion to any order desired, which should enable the hypothesis in presented in \cite{SOO} to be tested for a large class of model-specific matter couplings.\par
\indent
We will first rewrite the nine equations stemming from the quantum constraints in terms of the deviation matrix $\epsilon_{ab}$, using the results from \cite{EYO}.  Section 2 puts the constraints into standard form amenable to the expansion (\ref{MINK1}), which is carried out in sections 3 and 4.  We then create a library of terms which can be utilized to examine more general models of interest.\par
\indent  
In section 5 we represent the constraints as a linear operator acting on $\epsilon_{ae}$ plus nonlinear corrections suppressed by a small dimensionless coupling constant $G\Lambda$.  In a matrix representation acting on a nine dimensional space of deviation matrix elements $\epsilon_{ae}$, this has the interpretation of the flow of model-specific matter sources under the action of a new type of nonlinear transformation.  The linear part of the transformation has the representation of a nine by nine matrix of functional integro-differential operators.  The differentiation occurs in the direction of spatial position $\boldsymbol{x}\in\Sigma$ and also in the 
direction $\Gamma$ of the functional space of quantum fields $A^a_i$.  It is convenient to think of these variables as comprising the base space of a fibre bundle, with $\epsilon_{ae}$ as a section on the bundle undergoing a nonlinear transformation.\par
\indent  
The solution to the quantum constraints then amounts to finding the fixed points of this transformation, using the matter contributions as a driving input.  A full solution requires inversion of the linear part of the transformation followed by iterative perturbation expansion.  In section 6 we suggest a method to carry out this inversion, in analogy to the solution of the Gauss' law constraint discussed in \cite{EYO}.\par
\indent  
The precise determination of the propagation kernels required for full inversion require solution of linear first-order and a linear second-order differential equations with nonconstant coefficients in the full theory, which we do not carry out in this paper.  In section 7 we illustrate the solution for the deviation matrix by nonlinear iteration of the aforementioned linear transformation.  
We also illustrate the construction of the generalized Kodama state for the full theory for gravity quantized with the Klein--Gordon scalar field, as an asymptotic expansion relative to the pure Kodama state in powers of $G\Lambda$.  The coefficients of the expansion involve propagation of the nine-dimensional `generalized' matter charges, both in position 
space $\Sigma$ and also in the direction of the functional space  of fields $\Gamma_A$.\par
\indent

\section{Standard form for $q_0$, $q_1$ and $q_2$ in Ashtekar connection variables}

\noindent
As a review from \cite{EYO}, the coefficients of singularity determined from the quantum Hamiltonian constraint are given by

\begin{eqnarray}
\label{REVIEW}
q_0=\hbox{det}B\bigl(Var\Psi+\Lambda\hbox{det}\Psi\bigr)+G\Omega_0=0;\nonumber\\
q_1={\partial \over {\partial{A}^a_i}}(\epsilon_{ijk}\epsilon^{abc}B^k_{c}\Psi_{be}B^j_{e})+\epsilon_{ijk}\epsilon^{abc}D^{kj}_{cb}(\Psi_{ae}B^i_{e})\nonumber\\
+{\Lambda \over 4}{\partial \over {\partial{A}^a_i}}(\epsilon_{ijk}\epsilon^{abc}\Psi_{ce}\Psi_{bf}B^k_{e}B^j_{f})+G\Omega_{1}=0;\nonumber\\
q_2={\Lambda \over 6}{\partial \over {\partial{A^a_i}}}{\partial \over {\partial{A^b_j}}}
(\epsilon_{ijk}\epsilon^{abc}\Psi_{ce}B^k_{e})+36=0.
\end{eqnarray}

\noindent
It will be useful to express the constraints (\ref{REVIEW}) in the form of an operator acting on the nine-dimensional vector space of CDJ matrix elements.  This requires that the constraints be put into standard form.  The kinematic constraints, being linear in the CDJ matrix, are already in standard form.  This leaves the terms of the quantum Hamiltonian constraint $q_0$, $q_1$ and $q_2$.

\subsection{Semiclassical  term $q_0$}

\noindent
Starting with the semiclassical term $H_{cl}=q_0$, which is of zeroth order in singularity, we have

\begin{eqnarray}
\label{HAA}
q_0=\hbox{det}B\bigl(Var\Psi+\Lambda\hbox{det}\Psi\bigr)+G\Omega_0.
\end{eqnarray}

\noindent
Note that the condition $q_0=0$ implies a well-defined solution only for nondegenerate magnetic fields $B^i_a$ when there is matter in the theory (e.g. $\Omega_0\neq{0}$).  The case of degenerate $B$ is not applicable to the generalized Kodama states.\par
\indent

\subsection{First-order singularity $q_1$}

The first-order singularity term, given by $q_1$ is a first-order nonlinear differential equation given by

\begin{eqnarray}
\label{QUETWO}
q_1={\partial \over {\partial{A}^a_i}}(\epsilon_{ijk}\epsilon^{abc}B^k_{c}\Psi_{be}B^j_{e})+\epsilon_{ijk}\epsilon^{abc}D^{kj}_{cb}(\Psi_{ae}B^i_{e})\nonumber\\
+{\Lambda \over 4}{\partial \over {\partial{A}^a_i}}(\epsilon_{ijk}\epsilon^{abc}\Psi_{ce}\Psi_{bf}B^k_{e}B^j_{f})+G\Omega_{1}.
\end{eqnarray}

\noindent
Equation (\ref{QUETWO}) can be thought of as the functional divergence of linear and quadratic CDJ matrix terms with respect to 2-index and 4-index $SU(2)_{-}$ tensorial indices, which resembles a minisuperspace equation but is actually for the full theory.  The matter contribution $\Omega_1$ is present for matter models with Hamiltonians quadratic in conjugate momenta.  The functional form of the solution is the same for all $x$ (dependence of which is suppressed), since the $x$ dependence must be frozen in order that the integration can be carried out in the direction of the functional space of fields $A^a_i,\phi$.  We now place the gravitational part into standard form.

\subsection{Linear parts of $q_1$}

Expanding first the part of (\ref{QUETWO}) linear in CDJ matrix elements, we have

\begin{eqnarray}
\label{QONE}
(q_1)_{linear}={\partial \over {\partial{A}^a_i}}(\epsilon_{ijk}\epsilon^{abc}B^j_{e}B^k_{c}\Psi_{be})
+\epsilon_{ijk}\epsilon^{abc}D^{kj}_{cb}(\Psi_{ae}B^i_{e})\nonumber\\
=\epsilon_{ijk}\epsilon^{abc}\bigl(D^{ji}_{ea}B^k_c+B^j_{e}D^{ki}_{ca}\bigr)\Psi_{be}
+\epsilon_{ijk}\epsilon^{abc}D^{kj}_{cb}B^i_{e}\Psi_{ae}
+\epsilon_{ijk}\epsilon^{abc}B^j_{e}B^k_{c}{\partial \over {\partial{A}^a_i}}\Psi_{be}
\end{eqnarray}

\noindent
Starting with the first term on the right hand side of (\ref{QONE}),

\begin{eqnarray}
\label{QTWO}
\epsilon_{ijk}\epsilon^{abc}D^{ji}_{ea}B^k_{c}\Psi_{be}
=\epsilon_{ijk}\epsilon^{abc}\epsilon^{jil}\epsilon_{ead}A^d_{l}B^k_{c}\Psi_{be}\nonumber\\
=(\epsilon_{ijk}\epsilon^{jil})(\epsilon^{abc}\epsilon_{ead})A^d_{l}B^k_{c}\Psi_{be}
=(-2\delta^l_{k})(-1)\bigl(\delta^b_{e}\delta^c_{d}-\delta^b_{d}\delta^c_{e}\bigr)A^d_{l}B^k_{c}\Psi_{be}\nonumber\\
=2\bigl(\delta^b_{e}\delta^c_{d}-\delta^b_{d}\delta^c_{e}\bigr)C^d_{c}\Psi_{be}
=2\bigl(\delta^b_{e}\hbox{tr}C-C^b_{e}\bigr)\Psi_{be}.
\end{eqnarray}

\noindent
Moving on to the second term on the right hand side of (\ref{QONE}),

\begin{eqnarray}
\label{QTHREE}
\epsilon_{ijk}\epsilon^{abc}B^j_{e}D^{ki}_{ca}\Psi_{be}
=\epsilon_{ijk}\epsilon^{abc}B^j_{e}\epsilon^{kil}\epsilon_{cad}A^d_{l}\Psi_{be}\nonumber\\
=(\epsilon_{ijk}\epsilon^{kil})(\epsilon^{abc}\epsilon_{cad})B^j_{e}A^d_{l}
=(2\delta^l_{j})(2\delta^b_{d})B^j_{e}A^d_{l}=4C^b_{e}\Psi_{be}
\end{eqnarray}

\noindent
Moving on to the third term on the right hand side of (\ref{QONE}),

\begin{eqnarray}
\label{QFOUR}
\epsilon_{ijk}\epsilon^{abc}D^{kj}_{cb}B^i_{e}\Psi_{ae}
=\epsilon_{ijk}\epsilon^{abc}\epsilon^{kjl}\epsilon_{cbd}A^d_{l}B^i_{e}\Psi_{ae}\nonumber\\
=(\epsilon_{ijk}\epsilon^{kjl})(\epsilon^{abc}\epsilon_{cbd})A^d_{l}B^i_{e}\Psi_{ae}
=(-2\delta^l_{i})(-2\delta^a_d)A^d_{l}B^i_{e}\Psi_{ae}=4C^a_{e}\Psi_{ae}=4C^b_{e}\Psi_{be}
\end{eqnarray}

\noindent
Finally, the fourth term on the right hand side of (\ref{QONE})

\begin{eqnarray}
\label{QFIVE}
\epsilon_{ijk}\epsilon^{abc}B^j_{e}B^k_{c}{\partial \over {\partial{A}^a_i}}\Psi_{be}
=(\hbox{det}B)\epsilon^{abc}\epsilon_{ecd}(B^{-1})^d_{i}{\partial \over {\partial{A}^a_i}}\Psi_{be}
\end{eqnarray}

\noindent
So the total contribution to $q_1$ linear in the CDJ matrix elements is given by the sum of (\ref{QTWO}), (\ref{QTHREE}), (\ref{QFOUR}) and (\ref{QFIVE}), or

\begin{eqnarray}
\label{QONETOT}
(q_1)_{linear}={\partial \over {\partial{A}^a_i}}(\epsilon_{ijk}\epsilon^{abc}B^j_{e}B^k_{c}\Psi_{be})
+\epsilon_{ijk}\epsilon^{abc}D^{kj}_{cb}(\Psi_{ae}B^i_{e})\nonumber\\
=(\hbox{det}B)\epsilon^{abc}\epsilon_{ecd}(B^{-1})^d_{i}{{\partial\Psi_{be}} \over {\partial{A}^a_i}}
+\bigl(2\bigl(\delta^b_{e}\hbox{tr}C-C^b_{e}\bigr)\Psi_{be}\bigr)
+(4C^b_{e}\Psi_{be})+(4C^b_{e}\Psi_{be})\nonumber\\
=\Bigl[(\hbox{det}B)\epsilon^{bac}\epsilon_{edc}(B^{-1})^d_{i}{\partial \over {\partial{A}^a_i}}+6C^b_{e}+2\delta^b_{e}\hbox{tr}C\Bigr]\Psi_{be}.
\end{eqnarray}

\subsection{Quadratic parts of $q_1$}

\indent
Moving on to the quadratic CDJ matrix contributions to $q_1$, we expand the part of (\ref{QUETWO}) quadratic in CDJ matrix elements.

\begin{eqnarray}
\label{QSIX}
(q_1)_{quad}={\Lambda \over 4}{\partial \over {\partial{A}^a_i}}(\epsilon_{ijk}\epsilon^{abc}\Psi_{ce}\Psi_{bf}B^k_{e}B^j_{f})\nonumber\\
={\Lambda \over 4}\epsilon_{ijk}\epsilon^{abc}\bigl(D^{ki}_{ea}B^j_{f}+B^k_{e}D^{ji}_{fa}\bigr)\Psi_{ce}\Psi_{bf}
+{\Lambda \over 4}\epsilon_{ijk}\epsilon^{abc}B^k_{e}B^j_{f}{\partial \over {\partial{A}^a_i}}\Psi_{ce}\Psi_{bf}
\end{eqnarray}

\noindent
Starting with the first term on the right hand side of (\ref{QSIX}),

\begin{eqnarray}
\label{QSEVEN}
{\Lambda \over 4}\epsilon_{ijk}\epsilon^{abc}D^{ki}_{ea}B^j_{f}\Psi_{ce}\Psi_{bf}
={\Lambda \over 4}\epsilon_{ijk}\epsilon^{abc}\epsilon^{kil}\epsilon_{ead}A^d_{l}B^j_{f}\Psi_{ce}\Psi_{bf}\nonumber\\
={\Lambda \over 4}(\epsilon_{ijk}\epsilon^{kil})(\epsilon^{abc}\epsilon_{ead})A^d_{l}B^j_{f}\Psi_{ce}\Psi_{bf}
={\Lambda \over 4}(2\delta^l_j)(-1)\bigl(\delta^b_{e}\delta^c_{d}-\delta^b_{d}\delta^c_{e}\bigr)A^d_{l}B^j_{f}
\Psi_{ce}\Psi_{bf}\nonumber\\
={\Lambda \over 4}(-2)\bigl(\delta^b_{e}\delta^c_{d}-\delta^b_{d}\delta^c_{e}\bigr)C^d_{f}\Psi_{ce}\Psi_{bf}
={\Lambda \over 4}(-2)\bigl(\delta^b_{e}C^c_{f}-\delta^c_{e}C^b_{f}\bigr)\Psi_{ce}\Psi_{bf}
\end{eqnarray}

\noindent
Moving on the the second term on the right hand side of (\ref{QSIX}),

\begin{eqnarray}
\label{QEIGHT}
{\Lambda \over 4}\epsilon_{ijk}\epsilon^{abc}B^k_{e}D^{ji}_{fa}\Psi_{ce}\Psi_{bf}
={\Lambda \over 4}\epsilon_{ijk}\epsilon^{abc}\epsilon^{jil}\epsilon_{fad}A^d_{l}B^k_{e}\Psi_{ce}\Psi_{bf}\nonumber\\
={\Lambda \over 4}(\epsilon_{ijk}\epsilon^{jil})(\epsilon^{abc}\epsilon_{fad})A^d_{l}B^k_{e}\Psi_{ce}\Psi_{bf}
={\Lambda \over 4}(-2\delta^l_k)(-1)\bigl(\delta^b_{f}\delta^c_{d}-\delta^b_{d}\delta^c_{f}\bigr)A^d_{l}B^k_{e}
\Psi_{ce}\Psi_{bf}\nonumber\\
={\Lambda \over 4}(2)\bigl(\delta^b_{f}\delta^c_{d}-\delta^b_{d}\delta^c_{f}\bigr)C^d_{e}\Psi_{ce}\Psi_{bf}
={\Lambda \over 4}(2)\bigl(\delta^b_{f}C^c_{e}-\delta^c_{f}C^b_{e}\bigr)\Psi_{ce}\Psi_{bf}
\end{eqnarray}

\noindent
And finally, the last term on the right hand side of (\ref{QSIX}),

\begin{eqnarray}
\label{QNINE}
{\Lambda \over 4}\epsilon_{ijk}\epsilon^{abc}B^k_{e}B^j_{f}{\partial \over {\partial{A}^a_i}}\Psi_{ce}\Psi_{bf}
={\Lambda \over 4}(\hbox{det}B)\epsilon^{abc}\epsilon_{fed}(B^{-1})^d_{i}{\partial \over {\partial{A}^a_i}}\Psi_{ce}\Psi_{bf}
\end{eqnarray}

\noindent
So the total contribution to $q_1$ quadratic in the CDJ matrix elements is given by the sum of (\ref{QSEVEN}), (\ref{QEIGHT}) and (\ref{QNINE}), or

\begin{eqnarray}
\label{QTEN}
(q_1)_{quad}={\Lambda \over 4}{\partial \over {\partial{A}^a_i}}(\epsilon_{ijk}\epsilon^{abc}\Psi_{ce}\Psi_{bf}B^k_{e}B^j_{f})
={\Lambda \over 4}(\hbox{det}B)\epsilon^{abc}\epsilon_{fed}(B^{-1})^d_{i}{\partial \over {\partial{A}^a_i}}\Psi_{ce}\Psi_{bf}
\nonumber\\
+\Bigl({\Lambda \over 4}(-2)\bigl(\delta^b_{e}C^c_{f}-\delta^c_{e}C^b_{f}\bigr)\Psi_{ce}\Psi_{bf}\Bigr)
+\Bigl({\Lambda \over 4}(2)\bigl(\delta^b_{f}C^c_{e}-\delta^c_{f}C^b_{e}\bigr)\Psi_{ce}\Psi_{bf}\Bigr)\nonumber\\
\end{eqnarray}

\noindent
It can be shown by reshuffling indices that the last two terms on the right hand side of (\ref{QTEN}) are equal, leading to the result that

\begin{eqnarray}
\label{QELEVEN}
(q_1)_{quad}=\Bigl[{\Lambda \over 4}(\hbox{det}B)\epsilon^{abc}\epsilon_{fed}(B^{-1})^d_{i}{\partial \over {\partial{A}^a_i}}
+\Lambda\bigl(\delta^b_{f}C^c_{e}-\delta^c_{f}C^b_{e}\bigr)\Bigr]\Psi_{ce}\Psi_{bf}
\end{eqnarray}

\subsection{Total contribution to functional divergence terms}

\noindent
So the total contribution to the term first-order in singularity is given by the sum of (\ref{QONETOT}) and (\ref{QELEVEN})

\begin{eqnarray}
\label{SINGLE}
q_{1}=\Bigl[(\hbox{det}B)\epsilon^{abc}\epsilon_{dec}(B^{-1})^d_{i}{\partial \over {\partial{A}^a_i}}+6C^b_{e}+2\delta^b_{e}\hbox{tr}C\Bigr]\Psi_{be}\nonumber\\
+{\Lambda \over 4}(\hbox{det}B)\epsilon^{abc}\epsilon_{dfe}(B^{-1})^d_{i}{\partial \over {\partial{A}^a_i}}
+\Lambda\bigl(\delta^b_{f}C^c_{e}-\delta^c_{f}C^b_{e}\bigr)\Bigr]\Psi_{ce}\Psi_{bf}+G\Omega_1
\end{eqnarray}

\subsection{Functional Laplacian term}

The functional Laplacian term $q_2$ is given by

\begin{eqnarray}
\label{QUTWO}
q_2={\Lambda \over 6}{\partial \over {\partial{A^a_i}}}{\partial \over {\partial{A^b_j}}}
(\epsilon_{ijk}\epsilon^{abc}\Psi_{ce}B^k_{e})+36=\Delta^{be}\Psi_{be}+36.
\end{eqnarray}

\noindent
This can be expanded as

\begin{eqnarray}
\label{QUTHREE}
q_2=\epsilon_{ijk}\epsilon^{abc}B^k_{e}{\partial \over {\partial{A^a_i}}}{\partial \over {\partial{A^b_j}}}\Psi_{ce}
+\epsilon_{ijk}\epsilon^{abc}D^{ki}_{ea}{\partial \over {\partial{A^b_j}}}\Psi_{ce}\nonumber\\
+\epsilon_{ijk}\epsilon^{abc}D^{kj}_{eb}{\partial \over {\partial{A^a_i}}}\Psi_{ce}
+\epsilon_{ijk}\epsilon^{abc}\Bigl({\partial \over {\partial{A^a_i}}}{\partial \over {\partial{A^b_j}}}B^k_{e}\Bigr)\Psi_{ce}+36
\end{eqnarray}

\noindent
(I) We now evaluate the contributions to $q_2$ first-order in derivatives.  Starting with the second term of the right hand side of (\ref{QUTWO}), 

\begin{eqnarray}
\label{HAMMUT}
\epsilon_{ijk}\epsilon^{abc}D^{ki}_{ea}{\partial \over {\partial{A}^b_j}}
=\epsilon_{ijk}\epsilon^{abc}\epsilon^{kil}\epsilon_{ead}(A^d_{l})
{\partial \over {\partial{A}^b_j}}
=(2\delta^l_{j})(-1)\bigl(\delta^b_{e}\delta^c_{d}-\delta^b_{d}\delta^c_{e}\bigr)A^d_{l}
{\partial \over {\partial{A}^b_j}}\nonumber\\
=-2\bigl(\delta^b_{e}\delta^c_{d}-\delta^b_{d}\delta^c_{e}\bigr)
\Bigl(A{\partial \over {\partial{A}}}\Bigr)^d_{b}
=-2\Bigl[\Bigl(A{\partial \over {\partial{A}}}\Bigr)^c_{e}-\delta^c_{e}\hbox{tr}
\Bigl(A{\partial \over {\partial{A}}}\Bigr)\Bigr].
\end{eqnarray}

\noindent
which is traceless.  The third terms on the right hand side of (\ref{QUTWO}) and the second term are both equal.\par
\indent  
Expanding the third term on the right hand side of (\ref{QUTWO}), 

\begin{eqnarray}
\label{HAMMUT1}
\epsilon_{ijk}\epsilon^{abc}D^{kj}_{eb}{\partial \over {\partial{A}^a_i}}
=\epsilon_{ijk}\epsilon^{abc}\epsilon^{kjl}\epsilon_{ebd}(A^d_{l})
{\partial \over {\partial{A}^b_j}}
=(-2\delta^l_{i})\bigl(\delta^a_{e}\delta^c_{d}-\delta^a_{d}\delta^c_{e}\bigr)A^d_{l}{\partial \over {\partial{A}^a_i}}\nonumber\\
=-2\bigl(\delta^a_{e}\delta^c_{d}-\delta^a_{d}\delta^c_{e}\bigr)\Bigl(A{\partial \over {\partial{A}}}\Bigr)^d_{a}
=-2\Bigl[\Bigl(A{\partial \over {\partial{A}}}\Bigr)^c_{e}-\delta^c_{e}\hbox{tr}\Bigl(A{\partial \over {\partial{A}}}\Bigr)\Bigr].
\end{eqnarray}

\noindent
which is equal to (\ref{HAMMUT}).\par
\noindent
(II) We now look at the contribution to $q_2$ of second order in derivatives.  The last term of  (\ref{QUTWO}) is given by

\begin{eqnarray}
\label{HAMMUT2}
\epsilon_{ijk}\epsilon^{abc}{\partial \over {\partial{A}^a_i}}{\partial \over {\partial{A}^b_j}}(B^k_{e})
=\epsilon_{ijk}\epsilon^{abc}{\partial \over {\partial{A}^a_i}}{\partial \over {\partial{A}^b_j}}\bigl(\epsilon^{klm}\partial_{l}A_{em}
+{1 \over 2}\epsilon^{klm}\epsilon_{edf}A^d_{l}A^f_{m}\bigr),
\end{eqnarray}

\noindent
where we have made use of the unconventional calculus, first used in \cite{EYO}, that functional and spatial differentiation must commute.  Continuing on,

\begin{eqnarray}
\label{HAMMMUTING}
\epsilon_{ijk}\epsilon^{abc}{\partial \over {\partial{A}^a_i}}{\partial \over {\partial{A}^b_j}}\bigl(\epsilon^{klm}\partial_{l}A_{em}
+{1 \over 2}\epsilon^{klm}\epsilon_{edf}A^d_{l}A^f_{m}\bigr)\nonumber\\
={1 \over 2}\epsilon_{ijk}\epsilon^{abc}\epsilon^{klm}\epsilon_{edf}
{\partial \over {\partial{A}^a_i}}
\bigl(\delta^d_{b}\delta^j_{l}A^f_{m}+A^d_{l}\delta^f_{b}\delta^j_{m}\bigr)\nonumber\\
={1 \over 2}\epsilon_{ijk}\epsilon^{abc}\epsilon^{klm}\epsilon_{edf}
\bigl(\delta^d_{b}\delta^j_{l}\delta^f_{a}\delta^i_{m}
+\delta^d_{a}\delta^i_{l}\delta^f_{b}\delta^j_{m}\bigr)\nonumber\\
={1 \over 2}\epsilon_{ijk}\epsilon^{abc}\epsilon^{kji}\epsilon_{eba}
+{1 \over 2}\epsilon_{ijk}\epsilon^{abc}\epsilon^{kij}\epsilon_{eab}=(-3)(-2\delta^c_{e})+3(2\delta^c_{e})=12\delta^c_{e}.
\end{eqnarray}

\noindent
which gives a total, upon combining (\ref{HAMMUT}), (\ref{HAMMUT1}), (\ref{HAMMUT2}), 
(\ref{HAMMMUTING}), of

\begin{equation}
\label{CUUUUE2}
q_2={\Lambda \over 6}\biggl[\epsilon_{ijk}\epsilon^{abc}B^k_{e}{\partial \over {\partial{A^a_i}}}{\partial \over {\partial{A^b_j}}}
+4\bigl(\delta^c_{e}A^a_{k}-\delta^a_{e}A^c_{k}\bigr){\partial \over {\partial{A^a_k}}}+12\delta^c_{e}\biggr]\Psi_{ce}+36,
\end{equation}

\section{Expansion of the quantized kinematic constraints relative to the pure Kodama state}
\par
\medskip
\indent
Now that the coefficients of the singularities for the quantum Hamiltonian constraint have been put into standard form, we next expand these coefficients relative to the pure Kodama state $\Psi_{Kod}$.  The parametrization of the CDJ matrix consists of the expansion

\begin{equation}
\label{ANSA}
\Psi_{ae}=-\bigl(\kappa^{-1}\delta_{ae}+\epsilon_{ae}\bigr)
\end{equation}

\noindent
where in (\ref{ANSA}) we have defined $\kappa=(1/6)\Lambda$\footnote{It will be convenient to use $\kappa$ instead of $\Lambda$ in many of the computations that follow, to avoid carrying around various numerical factors of $6$.}, and where the deviation matrix $\epsilon_{ae}$ consists of nine independent elements.  The elements $\epsilon_{ae}$ are independent of each other until a solution to the quantum constraints is found, whereupon its degrees of freedom become saturated.\par
\indent  
Let us begin with the kinematic constraints, which are simultaneously satisfied at the semiclassical and at the quantum levels (SQC) due to being linear in conjugate 
momenta \cite{TQFT4}, and hence linear in the CDJ matrix.  Note that the mass dimensions are 
$[\kappa^{-1}]=[\epsilon_{ae}]=-2$.  Starting with the quantized diffeomorphism constraint \cite{EYO},

\begin{eqnarray}
\label{DI}
\epsilon_{dae}\Psi_{ae}=\epsilon_{dae}\bigl[-\kappa^{-1}\delta_{ae}-\epsilon_{ae}\bigr]=G\widetilde{\tau}_{0d};\nonumber\\
\epsilon_{dae}\epsilon_{ae}=-G\widetilde{\tau}_{0d}.
\end{eqnarray}

\noindent
The antisymmetric elements of the CDJ deviation matrix $\epsilon_{ae}$ are determined by the local matter momentum $H_i$.  Since the constraint must be identically satisfied at all points ${x}$ in the spatial hypersurface $\Sigma_t$, labeled by time $t$, these elements are uniquely globally determined in the functional dependence upon the fields $(A^a_i(x),\phi^{\alpha}(x))$ indicated at all points.  Note that since this is a linear constraint it can be solved explicitly when the Ashtekar magnetic field $B^i_a$ is nondegenerate (invertible).  This invertibility condition is a direct consequence of the presence of local matter fields in the theory \cite{THIE2}.  Also, the pure Kodama part of the equation has vanished.  A final note for checks of consistency in the last line of 
(\ref{DI}) is that the mass dimension of $H_i$ is $[H_i]=[\phi]+[\partial]+[\pi]=1+1+2=4$, causing the mass dimensions to balance.\par
\indent
Now for the quantized Gauss' law constraint,

\begin{eqnarray}
\label{GEE}
B^i_{e}D_{i}\Psi_{ae}+GQ_a=B^i_{e}D_{i}\bigl[-\kappa^{-1}\delta_{ae}-\epsilon_{ae}\bigr]+GQ_a=0\nonumber\\
{B}^i_{e}D_{i}\delta_{ae}+B^i_{e}D_{i}\epsilon_{ae}=GQ_a\longrightarrow0+{B}^i_{e}D_{i}\epsilon_{ae}=GQ_a
\end{eqnarray} 

\noindent
Note again that the pure Kodama contribution has dropped out, leaving the remaining parts $\epsilon_{ae}$ to be determined by the matter fields.  This leads to three first-order differential equations in nine unknowns \cite{EYO}

\begin{equation}
\label{GOOS}
{B}^i_{e}D_{i}\epsilon_{ae}=\Bigl(\delta_{af}{{\partial} \over {\partial{t^g}}}
+C_{a}^{fg}\Bigr)\epsilon_{fg}=0=GQ_a
\end{equation}

\noindent
where $C\equiv{C^e_b}=A^e_{i}B^i_b$ is a $SU(2)\otimes{SU(2)}$-valued object which acts as a connection for parallel transport along the directions $t^e$, where $e=1,2,3$ represent 
SU(2) indices.  The integration of this partial differential equation requires as boundary conditions the specification of the values of three elements $\epsilon_{ae}$ on 
three linearly-independent 2-dimensional hypersurfaces dual to the direction of path-ordered integration of the partial differential equation (\ref{GOOS}) and the value of all 
elements not being solved for via this constraint (six) everywhere in $\Sigma$.  Three elements of $\epsilon_{ae}$ are already fixed by (\ref{DI}).  So let us decompose the CDJ 
matrix into its symmetric and its antisymmetric parts

\begin{equation}
\Psi_{ae}=\psi_{ae}+\epsilon_{aed}\psi_d
\end{equation}

\noindent
Then Gauss' law constraint can be solved as outlined in \cite{EYO} for three of the symmetric elements $\psi_{ae}$ in terms of the antisymmetric elements $\psi_d$, which are 
uniquely determined by the local matter momentum, and the remaining three symmetric elements.  This yields

\begin{eqnarray}
\label{SIMP}
B^i_{e}D_{i}\bigl(\psi_{ae}+\epsilon_{aed}\psi_d\bigr)=GQ_a\nonumber\\
B^i_{e}D_{i}\psi_{ae}=GQ_{a}-GB^i_{e}D_{i}\Bigl(\epsilon_{aed}{{B^j_{d}H_j} \over {\hbox{det}B}}\Bigr).
\end{eqnarray}

\noindent
So the Gauss' law constraint has been reduced via (\ref{SIMP}) from nine to three equations in six unknowns.  These equations can be solved explicitly for three elements in terms of the remaining three, which are determined by the Hamiltonian constraint.  Lastly, regarding mass dimensions, the parameters of path-ordered integration have $[t^e=[\partial]+[B]=1+2=3$ and 
the $SU(2)_{-}$ charge has $[Q]=[\phi]+[\pi]=1+2=3$, causing the mass dimensions on both sides of (\ref{GEE}) to balance.  In this work we will not use the diagonal CDJ matrix elements to solve the Gauss' law constraint as shown in \cite{EYO}, but rather the upper off-diagonal 
elements $\Psi_{12},\Psi_{23},\Psi_{31}$.

\section{Expansion of the Quantized Hamiltonian constraint relative to the pure Kodama state}
\par
\medskip

\noindent
We will now expand the Hamiltonian constraint in terms of the Ansatz (\ref{MINK1}).  This is not a linearization, since we will carry out the expansion to all orders.  The polynomial nature of the constraints in Ashtekar variables ensures termination of the series expansion in $\Lambda$ unlike in metric variables where the corresponding series is infinite.  The standard form of the Hamiltonian contribution to the constraints is then given, rewriting (\ref{SINGLE}) and (\ref{QTEN}) for completeness, by

\begin{eqnarray}
q_0=\hbox{det}B\bigl(Var\Psi+\Lambda\hbox{det}\Psi\bigr)+G\Omega_0
\end{eqnarray}

\begin{eqnarray}
\label{QUE2}
q_{1}=\Bigl[(\hbox{det}B)\epsilon^{abc}\epsilon_{dec}(B^{-1})^d_{i}{\partial \over {\partial{A}^a_i}}+6C^b_{e}+2\delta^b_{e}\hbox{tr}C\Bigr]\Psi_{be}\nonumber\\
+{\Lambda \over 4}(\hbox{det}B)\epsilon^{abc}\epsilon_{dfe}(B^{-1})^d_{i}{\partial \over {\partial{A}^a_i}}
+\Lambda\bigl(\delta^b_{f}C^c_{e}-\delta^c_{f}C^b_{e}\bigr)\Bigr]\Psi_{ce}\Psi_{bf}+G\Omega_1
\end{eqnarray}

\begin{equation}
\label{CUUUUE2}
q_2={\Lambda \over 6}\biggl[\epsilon_{ijk}\epsilon^{abc}B^k_{e}{\partial \over {\partial{A^a_i}}}{\partial \over {\partial{A^b_j}}}
+4\bigl(\delta^c_{e}A^a_{k}-\delta^a_{e}A^c_{k}\bigr){\partial \over {\partial{A^a_k}}}+12\delta^c_{e}\biggr]\Psi_{ce}+36,
\end{equation}

\subsection{Expansion of $q_2$ relative to the pure Kodama state}
\par
\medskip
\indent
We now expand the CDJ matrix about the isotropic pure Kodama solution in the quantized Hamiltonian constraint.  

\begin{eqnarray}
\label{CUUUUET}
q_2={\Lambda \over 6}\Bigl[\epsilon_{ijk}\epsilon^{abc}B^i_{e}{\partial \over {\partial{A}^b_j}}
{\partial \over {\partial{A}^c_k}}
+4\bigl(\delta^c_{e}A^a_{k}-\delta^a_{e}A^c_{k}\bigr){\partial \over {\partial{A^a_k}}}+12\delta_{ae}\Bigr]\bigl(-{6 \over \Lambda}\delta_{ae}-\epsilon_{ae}\bigr)+36\nonumber\\
={\Lambda \over 6}\bigl(-{6 \over \Lambda}\delta_{ae}\bigr)(12\delta_{ae})+36+{\Lambda \over 6}\Delta_{ae}\epsilon_{ae}={\Lambda \over 6}\Delta_{ae}\epsilon_{ae}
\end{eqnarray}

\noindent
The pure Kodama contribution has cancelled out the inhomogeneous factor of $36$.  Equation  
(\ref{CUUUUET}) can be thought of as a set of nine individual `generalized' functional Laplacian operators $\hat{\Delta}_{ae}$, each acting on an individual element of $\epsilon_{ae}$.  The functional Laplacians are given by

\begin{eqnarray}
\label{FUNCLAP}
\Delta_{ae}=\epsilon_{ijk}\epsilon^{abc}B^i_{e}{\partial \over {\partial{A}^b_j}}
{\partial \over {\partial{A}^c_k}}
+4\bigl(\delta^a_{e}A^c_{k}-\delta^c_{e}A^a_{k}\bigr){\partial \over {\partial{A^a_k}}}+12\delta_{ae}
\end{eqnarray}

\noindent
Since each elements of $\Delta_{ae}$ is at worst a linear record-order differential operator at each point $x$ with respect to the functional space of fields, its inversion can be carried out in principle by heat kernel methods and existence theorems regarding the solutions can be applied.  We relegate this technical step of the full superspace theory to future publication.

\subsection{Expansion of the nonderivative parts of $q_1$ relative to the pure Kodama state}
\par
\medskip
\indent
We are now ready to carry out the expansion on the nonderivative parts of $q_1$,

\begin{equation}
\Psi_{ae}=-\bigl({1 \over \kappa}\delta_{ae}+\epsilon_{ae}\bigr)
\end{equation}

\noindent
where $\kappa=(1/6)\Lambda$.  Starting with the nonderivative linear term in $q_1$in the CDJ matrix,

\begin{eqnarray}
\label{NONNDER}
\bigl(6C^b_{f}+2\delta^b_{f}\hbox{tr}C)\Psi_{ae}=\bigl(6C^b_{f}+2\delta^b_{f}\hbox{tr}C)\bigl(-{1 \over \kappa}\delta_{bf}-\epsilon_{bf}\bigr)\nonumber\\
=-{1 \over \kappa}\bigl(6C^b_{f}\delta_{bf}+2\delta^b_{f}\delta_{bf}\hbox{tr}C)-\bigl(6C^b_{f}+2\delta^b_{f}\hbox{tr}C\bigr)\epsilon_{bf}\nonumber\\
=-\Bigl({{12} \over \kappa}\Bigr)\hbox{tr}C-\bigl(6C^b_{f}+2\delta^b_{f}\hbox{tr}C\bigr)\epsilon_{bf}.
\end{eqnarray}

\noindent
Now the for the nonderivative part of $q_1$ quadratic in the CDJ matrix,

\begin{eqnarray}
\label{NONNDER1}
6\kappa\bigl(\delta^a_eC^b_f-\delta^b_eC^a_f\bigr)\Psi_{bf}=6\kappa\bigl(\delta^a_{e}\delta^b_{d}-\delta^b_{e}\delta^a_{d}\bigr)C^d_{f}\Psi_{ae}\Psi_{bf}\nonumber\\
=6\kappa{C}^{ab}_{ef}\bigl({1 \over \kappa}\delta_{ae}+\epsilon_{ae}\bigr)\bigl({1 \over \kappa}\delta_{bf}+\epsilon_{bf}\bigr)\nonumber\\
=6\kappa{C}^{ab}_{ef}\Bigl({1 \over {\kappa^2}}\delta_{ae}\delta_{bf}+{1 \over \kappa}\delta_{ae}\epsilon_{bf}
+{1 \over \kappa}\epsilon_{ae}\delta_{bf}+\epsilon_{ae}\epsilon_{bf}\Bigr).
\end{eqnarray}

\noindent
Concentrating on the first term of (\ref{NONNDER1}), the contribution due to the pure Kodama state is given by

\begin{eqnarray}
\label{NONNDER2}
6\kappa{C}^{ab}_{ef}{1 \over {\kappa^2}}\delta_{ae}\delta_{bf}
={6 \over \kappa}\bigl(\delta^a_{e}\delta^b_{d}-\delta^b_{e}\delta^a_{d}\bigr)C^d_{f}(\delta_{ae}\delta_{bf})\nonumber\\
={6 \over \kappa}\bigl(3\hbox{tr}C-\delta_{df}C^d_{f}\bigr)={{12\hbox{tr}C} \over \kappa}
\end{eqnarray}

\noindent
Note that the factor of 12 on the last line of (\ref{NONNDER2}) is precisely that needed to cancel the leading order term of (\ref{NONNDER}).\par
\indent  
Next we move on to the terms in (\ref{NONNDER1}) linear in $\epsilon_{ae}$.  Contrary to the inclination to double this term, thus avoiding its explicit calculation, let us proceed 
with its computation.

\begin{eqnarray}
\label{NONNDER3}
6\kappa{C}^{ab}_{ef}{1 \over \kappa}\delta_{ae}\epsilon_{bf}
=6\bigl(\delta^a_{e}\delta^b_{d}-\delta^b_{e}\delta^a_{d}\bigr)C^d_{f}(\delta_{ae}\epsilon_{bf})\nonumber\\
=6\bigl(3C^b_{f}-C^b_{f}\bigr)\epsilon_{bf}=12\hbox{tr}(C\epsilon)
\end{eqnarray}

\noindent
Moving on to the next linear term of (\ref{NONNDER1}),

\begin{eqnarray}
\label{NONNDER4}
6\kappa{C}^{ab}_{ef}{1 \over \kappa}\epsilon_{ae}\delta_{bf}
=6\bigl(\delta^a_{e}\delta^b_{d}-\delta^b_{e}\delta^a_{d}\bigr)C^d_{f}(\epsilon_{ae}\delta_{bf})\nonumber\\
=6\bigl((\hbox{tr}C)\hbox{tr}\epsilon-\epsilon_{ab}\delta^a_{d}C^d_{f}\delta_{bf}\bigr)
=6(\hbox{tr}C)\hbox{tr}\epsilon-6\hbox{tr}(C\epsilon)
\end{eqnarray}

\noindent
(\ref{NONNDER4}) is not quite the same as the previous term (\ref{NONNDER3}), which may be counter to any preconceived intuition on symmetries.  This is a manifestation of a 
property which has a statistical interpretation, known as skewness.\par
\indent  
Lastly, there is a term of (\ref{NONNDER1}) quadratic in $\epsilon$

\begin{equation}
\label{NONNDER5}
6\kappa{C}^{ab}_{ef}\epsilon_{ae}\epsilon_{bf}=\bigl(\delta^a_eC^b_f-\delta^b_eC^a_f\bigr)\epsilon_{ae}\epsilon_{bf}={\Lambda}C^{ab}_{ef}\epsilon_{ae}\epsilon_{bf}.
\end{equation}

\noindent
This is the first term featuring a departure from linearity.  Notice that this term is suppressed by a factor of order $\sim\Lambda$ which is small, therefore one would expect a linearized treatment of this particular term (\ref{NONNDER5}) to be extremely accurate.\par
\indent
To recapitulate, combining all contributions, the nonderivative part of the expansion of $q_1$ is given, combining (\ref{NONNDER}), (\ref{NONNDER2}), 
(\ref{NONNDER4}) and (\ref{NONNDER5}), by

\begin{eqnarray}
\label{NONNDER6}
\bigl(6C^b_{f}+2\delta^b_{f}\hbox{tr}C\bigr)\Psi_{bf}+6\kappa{C}^{ba}_{fe}\Psi_{ae}\Psi_{bf}
=-{{12\hbox{tr}C} \over \kappa}-\bigl(6C^b_{f}+2\delta^b_{f}\hbox{tr}C\bigr)\epsilon_{bf}\nonumber\\
+{{12\hbox{tr}C} \over \kappa}+12\hbox{tr}(C\epsilon)+6(\hbox{tr}\epsilon)(\hbox{tr}C)-6\hbox{tr}(\epsilon{C})
+6\kappa{C}^{ba}_{fe}\epsilon_{ae}\epsilon_{bf}\nonumber\\
=-\bigl(6C^b_{f}+2\delta^b_{f}\hbox{tr}C\bigr)\epsilon_{bf}
+6(\hbox{tr}\epsilon)(\hbox{tr}C)+6\hbox{tr}(\epsilon{C})+6\kappa{C}^{ba}_{fe}\epsilon_{ae}\epsilon_{bf}\nonumber\\
=6\hbox{tr}(\epsilon{C})+6(\hbox{tr}\epsilon)(\hbox{tr}C)-\bigl(6C^b_{f}+2\delta^b_{f}\hbox{tr}C\bigr)\epsilon_{bf}+{\Lambda}C^{ba}_{fe}\epsilon_{ae}\epsilon_{bf},
\end{eqnarray}

\noindent
where we have used the cyclic property of the trace.

\subsection{Expansion of the functional derivative parts of $q_1$ relative to the pure Kodama state}
\par
\medskip
\indent
Under the Ansatz

\begin{equation}
\Psi_{bf}=-\bigl({1 \over \kappa}\delta_{bf}+\epsilon_{bf}\bigr)
\end{equation}

\noindent
and using (\ref{SINGLE}), we must first focus on the gravitational part of 

\begin{eqnarray}
\label{SINGLE}
\Bigl[(\hbox{det}B)\epsilon^{abc}\epsilon_{ecd}(B^{-1})^d_{i}{{\partial\Psi_{be}} \over {\partial{A}^a_i}}
+{\Lambda \over 4}(\hbox{det}B)\epsilon^{abc}\epsilon_{fed}(B^{-1})^d_{i}{\partial \over {\partial{A}^a_i}}
+\Lambda\bigl(\delta^b_{f}C^c_{e}-\delta^c_{f}C^b_{e}\bigr)\Psi_{ce}\Psi_{bf}\Bigr]+G\Omega_1.
\end{eqnarray}

\noindent
This gives

\begin{eqnarray}
\label{FUNCTIONAL}
\epsilon^{abc}(\hbox{det}B)(B^{-1})^d_{i}{\partial \over {\partial{A}^a_i}}
\Bigl[\epsilon_{ecd}\Psi_{be}+{3 \over 2}\kappa\epsilon_{fed}\Psi_{ce}\Psi_{bf}\Bigr].
\end{eqnarray}

\noindent
Starting with the linear term of (\ref{FUNCTIONAL}), we have

\begin{eqnarray}
\label{FUNCTIONAL1}
\epsilon^{abc}(\hbox{det}B)(B^{-1})^d_{i}{\partial \over {\partial{A}^a_i}}\epsilon_{ecd}\bigl(-{1 \over \kappa}-\epsilon_{be}\bigr)
=-(\hbox{det}B)\epsilon^{abc}\epsilon_{ecd}(B^{-1})^d_{i}{\partial \over {\partial{A}^a_i}}\epsilon_{be}\nonumber\\
=(\hbox{det}B)\epsilon^{abc}\epsilon_{edc}(B^{-1})^d_{i}{\partial \over {\partial{A}^a_i}}\epsilon_{be}
=(\hbox{det}B)\bigl(\delta^a_e\delta^b_d-\delta^a_d\delta^b_e\bigr)(B^{-1})^d_{i}{\partial \over {\partial{A}^a_i}}\epsilon_{be}\nonumber\\
=(\hbox{det}B)\Bigl[(B^{-1})^b_i{\partial \over {\partial{A}^e_i}}-\delta^b_e\hbox{tr}(B^{-1}){\partial \over {\partial{A}}}\Bigr]\epsilon_{be}.
\end{eqnarray}

\noindent
Moving on to the quadratic term of (\ref{FUNCTIONAL}), we have

\begin{eqnarray}
\label{FUNCTIONAL2}
{3 \over 2}\kappa\epsilon^{abc}(\hbox{det}B)(B^{-1})^d_{i}{\partial \over {\partial{A}^a_i}}\epsilon_{fed}\Psi_{ce}\Psi_{bf}
={3 \over 2}\kappa(\hbox{det}B)\epsilon^{abc}\epsilon_{fed}{\partial \over {\partial{A}^a_i}}
\Bigl({1 \over \kappa}\delta_{ce}+\epsilon_{ce}\Bigr)\Bigl({1 \over \kappa}\delta_{bf}+\epsilon_{bf}\Bigr)\nonumber\\
={3 \over 2}\kappa(\hbox{det}B)\epsilon^{abc}\epsilon_{fed}{\partial \over {\partial{A}^a_i}}
\Bigl[{1 \over {\kappa^2}}\delta_{ce}\delta_{bf}
+{1 \over \kappa}\delta_{ce}\epsilon_{bf}+{1 \over \kappa}\epsilon_{ce}\delta_{bf}+\epsilon_{ce}\epsilon_{bf}\Bigr]\nonumber\\
=0+{3 \over 2}(\hbox{det}B)\epsilon^{abe}\epsilon_{fed}(B^{-1})^d_i{\partial \over {\partial{A}^a_i}}\epsilon_{bf}
+{3 \over 2}(\hbox{det}B)\epsilon^{abc}\epsilon_{bed}(B^{-1})^d_i{\partial \over {\partial{A}^a_i}}\epsilon_{ce}
+{3 \over 2}\kappa(\hbox{det}B)\epsilon^{abc}\epsilon_{fed}{\partial \over {\partial{A}^a_i}}\epsilon_{ce}\epsilon_{bf}
\end{eqnarray}

\noindent
Continuing with the expansion,

\begin{eqnarray}
\label{FUNCTIONAL3}
{3 \over 2}\kappa\epsilon^{abc}(\hbox{det}B)(B^{-1})^d_{i}{\partial \over {\partial{A}^a_i}}\epsilon_{fed}\Psi_{ce}\Psi_{bf}=
-{3 \over 2}(\hbox{det}B)\bigl(\delta^a_f\delta^b_d-\delta^a_d\delta^b_f\bigr)(B^{-1})^d_i{\partial \over {\partial{A}^a_i}}\epsilon_{bf}\nonumber\\
-{3 \over 2}(\hbox{det}B)\bigl(\delta^a_e\delta^c_d-\delta^a_d\delta^c_e\bigr)(B^{-1})^d_i{\partial \over {\partial{A}^a_i}}\epsilon_{ae}
+{3 \over 2}\kappa(\hbox{det}B)(B^{-1})^d_i{\partial \over {\partial{A}^a_i}}\epsilon^{abc}\epsilon_{dfe}\epsilon_{ce}\epsilon_{bf}\nonumber\\
=-{3 \over 2}(\hbox{det}B)\Bigl[(B^{-1})^b_i{\partial \over {\partial{A}^f_i}}-\delta^b_f\hbox{tr}(B^{-1}){\partial \over {\partial{A}}}\Bigr]\epsilon_{bf}
-{3 \over 2}(\hbox{det}B)\Bigl[(B^{-1})^c_i{\partial \over {\partial{A}^e_i}}-\delta^c_e\hbox{tr}(B^{-1}){\partial \over {\partial{A}}}\Bigr]\epsilon_{ce}\nonumber\\
+{3 \over 2}\kappa(\hbox{det}B)(B^{-1})^d_i{\partial \over {\partial{A}^a_i}}\epsilon^{abc}\epsilon_{dfe}\epsilon_{ce}\epsilon_{bf}.
\end{eqnarray}

\noindent
Note that the first two terms on the right hand side of (\ref{FUNCTIONAL3}) are equal to each other, and equal modulo a factor of $3/2$ to (\ref{FUNCTIONAL1}).  Combining (\ref{FUNCTIONAL}) and (\ref{FUNCTIONAL3}) produces a contribution due to the derivative terms of

\begin{eqnarray}
\label{FUNCTIONAL4}
2(\hbox{det}B)\Bigl[\delta^b_f\hbox{tr}(B^{-1}){\partial \over {\partial{A}}}-(B^{-1})^b_i{\partial \over {\partial{A}^f_i}}\Bigr]\epsilon_{bf}
+{3 \over 2}\kappa(\hbox{det}B)(B^{-1})^d_i{\partial \over {\partial{A}^a_i}}\epsilon^{abc}\epsilon_{dfe}\epsilon_{ce}\epsilon_{bf}.
\end{eqnarray}

\noindent
It is convenient to define a new vector field antisymmetric in indices $bc$ and antisymmetric in indices $fe$ given by

\begin{eqnarray}
\label{VECTOR}
{\partial \over {\partial{X}^{bc}_{fe}}}=
(B^{-1})^d_i{\partial \over {\partial{A}^a_i}}\epsilon^{abc}\epsilon_{dfe}
\end{eqnarray}

\noindent
and its single and double traces, given by

\begin{eqnarray}
\label{VECTOR1}
{\partial \over {\partial{X}^{b}_f}}=\sum_{ce}{\partial \over {\partial{X}^{bc}_{fe}}}
\equiv{\partial \over {\partial{X}^{bf}}};~~
{\partial \over {\partial{X}}}=\sum_{bf}{\partial \over {\partial{X}^{bf}}}.
\end{eqnarray}

\noindent
Equations (\ref{VECTOR1}) can as well be written in the form

\begin{eqnarray}
\label{VECTOR2}
{\partial \over {\partial{X}^{bf}}}=
\delta^b_f\hbox{tr}(B^{-1}){\partial \over {\partial{A}}}-(B^{-1})^b_i{\partial \over {\partial{A}^f_i}};~~
{\partial \over {\partial{X}}}=2\hbox{tr}(B^{-1}){\partial \over {\partial{A}}}
\end{eqnarray}

\noindent
The functional divergence is the divergence in the space of functions at each spatial point $\boldsymbol{x}$ in $\Sigma$, which has the interpretation of the next influx or outflux of a physical property indigenous to the CDJ matrix elements within a `volume' in the space of functions.  The question is what is the source that drives this flux relative to the $SU(2)$ directions.  One can see that it has to do with a certain combination of matter fields and their first-order quantum fluctuations encoded 
in $\Omega_1$.  Note that there is a trace only in $SU(2)_{-}$ indices in this divergence and not in position label $\boldsymbol{x}$.
Hence the divergence reduces one from nine degrees to freedom to one degrees of freedom per point for $\epsilon_{ae}$, and from 81 superficial to one degrees of freedom for 
$\epsilon_{ae}\epsilon_{bf}$.\par
\indent
Thus the total contribution to $q_1$ is given, combining the results of (\ref{NONNDER6}) and (\ref{FUNCTIONAL4}), by

\begin{eqnarray}
\label{DIVER}
2(\hbox{det}B){\partial \over {\partial{X^{bf}}}}\epsilon_{bf}+{3 \over 2}\kappa(\hbox{det}B){\partial \over {\partial{X}^{ab}_{ef}}}(\epsilon_{ae}\epsilon_{bf})
-\bigl(6C^b_{f}+2\delta^b_{f}\hbox{tr}C\bigr)\epsilon_{bf}\nonumber\\
+6(\hbox{tr}\epsilon)(\hbox{tr}C)+6\hbox{tr}(\epsilon{C})+6\kappa{C}^{ab}_{ef}\epsilon_{ae}\epsilon_{bf}+G\Omega_1
\end{eqnarray}

\noindent
One may in general desire to be able to write this in the form of a functional covariant derivative, but first let us simplify the linear part of the nonderivative terms of (\ref{DIVER}) involving $C$ in terms of $\epsilon_{bf}$.

\begin{equation}
-\bigl(6C^b_{f}+2\delta^b_{f}\hbox{tr}C\bigr)\epsilon_{bf}
+6(\delta_{bf}\hbox{tr}C)\epsilon_{bf}+6C^b_{f}\epsilon_{bf}=4\hbox{tr}C\hbox{tr}\epsilon
\end{equation}

\noindent
So, the overall contribution to $q_1$ due to quantum gravitational fluctuations about the pure Kodama state is given by

\begin{equation}
\label{DIVERGEN}
2\Bigl((\hbox{det}B){\partial \over {\partial{X}^{bf}}}+2\delta_{bf}\hbox{tr}C\Bigr)\epsilon_{bf}
+{3 \over 2}\kappa\Bigl((\hbox{det}B){\partial \over {\partial{X}^{ab}_{ef}}}+4C^{ab}_{ef}\Bigr)(\epsilon_{ae}\epsilon_{bf})+G\Omega_1
\end{equation}

\noindent
Equation (\ref{DIVERGEN}) is the total gravitational functional divergence.  Note that each term acts as a covariant derivative along a particular direction in function space, 
using $C^e_b=A^e_{i}B^i_b$ as a connection.  Recall that this is the same connection utilized in the $SU(2)$ representation of the Gauss' law constraint \cite{EYO}.  So the connection 
 gives the rule for parallel transport of the CDJ matrix elements both in functional and in position space, two mutually orthogonal sets of directions in the 
trivializable base space $(\Sigma,\Gamma)$ of the fibre bundle structure.  It will be convenient to use $\nabla$ for the covariant form of the functional divergence operator $\partial$ in order to place these derivatives on an equal footing as spatial gradients, highlighting the trivializability of the bundle.  In this notation we have, for the total contribution to $q_1$,

\begin{equation}
\label{CONNECTION}
q_1=\bigl(2(\hbox{det}B)\partial^{bf}+4\delta^f_{b}\hbox{tr}C\bigr)\epsilon_{bf}+{\Lambda \over 4}\bigl((\hbox{det}B)\partial^{ef}_{ab}
+4{C}^{ab}_{ef}\bigr)\epsilon_{ae}\epsilon_{bf}+G\Omega_1.
\end{equation}

\section{Expansion of $q_0$ relative to the pure Kodama state}
\par
\medskip
\indent
This is the only algebraic constraint to be solved, apart from the diffeomorphism constraint, and is nonlinear.  We will need to make use of a few identities.  Starting with the
Ansatz 

\begin{equation}
\Psi_{ae}=-\Bigl({1 \over \kappa}\delta_{ae}+\epsilon_{ae}\Bigr)
\end{equation}

\noindent
it suffices to calculate the $SU(2)$ invariants of the CDJ matrix.  We first calculate the trace and variance

\begin{eqnarray}
\hbox{tr}\Psi=-\Bigl({3 \over \kappa}+\hbox{tr}\epsilon\Bigr)\nonumber\\
{Var}\Psi=(\hbox{tr}\Psi)^{2}-\hbox{tr}\Psi^2\nonumber\\
=\Bigl[{3 \over \kappa}+\hbox{tr}\epsilon\Bigr]^2-\Bigl({1 \over \kappa}\delta_{ab}+\epsilon_{ab}\Bigr)({1 \over \kappa}\delta_{ba}+\epsilon_{ba}\Bigr)\nonumber\\
={9 \over {\kappa^2}}+{6 \over \kappa}\hbox{tr}\epsilon+(\hbox{tr}\epsilon)^2-
\Bigl({3 \over {\kappa^2}}+{2 \over \kappa}\hbox{tr}\epsilon+\hbox{tr}\epsilon^2\Bigr)\nonumber\\
={6 \over {\kappa^2}}+{4 \over \kappa}\hbox{tr}\epsilon+(\hbox{tr}\epsilon)^2-\hbox{tr}\epsilon^2\nonumber\\
\longrightarrow{Var}\Psi=Var{\epsilon}+{6 \over {\kappa^2}}+{4 \over \kappa}\hbox{tr}\epsilon
\end{eqnarray}

\noindent
The $SU(2)$ trace and variance have a statistical interpretation.\par
\noindent
(i)  If the values of the CDJ matrix were correlated to a normal distribution, the $SU(2)$ trace would correspond to the mean and the $SU(2)$ variance would correspond to the variance of statistical fluctuations about the mean.  But what is the interpretation of statistical fluctuations in
$SU(2)$?\par
\noindent  
(ii) We interpret these to be actual quantum gravitational fluctuations of order $\kappa\propto\Lambda$, due to the presence of matter fields.  These statistical moments would 
not exist for the pure Kodama state for which $\epsilon_{ab}=0$.  The pure Kodama state can therefore can be viewed in terms of normal statistical distributions as a spike of zero width.  Therefore
we expect that the presence of matter in a universe with a cosmological constant should induce quantum effects which should be distinguishable by our model.\par
\noindent
(iii) There is a large number of different matter models and combinations thereof that can be examined.  But whatever the model, the induced quantum effects should in principle fix the CDJ matrix elements.\par
\noindent
(iv) There is another statistical measure $\hbox{det}\epsilon$, the third-moment, which appears in our model.  In statistical language, it is a measure of the skewness of the
distribution.  So we need to determine what the analog for skewness  is for $SU(2)$.  Neverthless, it is a quantum gravitational effect which is carried over into the
semiclassical level, albeit suppresed by order $\kappa^2$ relative to the `statistical mean' and is present due to the cosmological constant.\par
\indent
This brings us to the final invariant that we will need in the expansion of the Hamiltonian constraint about the pure Kodama state, namely the determinant

\begin{equation}
\hbox{det}\Psi={1 \over 6}\epsilon_{abc}\epsilon_{def}\Psi_{ad}\Psi_{be}\Psi_{cf}
\end{equation}

\noindent
Substituting in the Ansatz,

\begin{eqnarray}
-6\hbox{det}\Psi
=\epsilon_{abc}\epsilon_{def}\bigl({1 \over \kappa}\delta_{ad}+\epsilon_{ad}\bigr)\bigl({1 \over \kappa}\delta_{be}+\epsilon_{be}\bigr)
\bigl({1 \over \kappa}\delta_{cf}+\epsilon_{cf}\bigr)\nonumber\\
=\kappa^{-3}\epsilon_{abc}\epsilon_{def}\delta_{ad}\delta_{be}\delta_{cf}
+3\kappa^{-2}\epsilon_{abc}\epsilon_{def}\delta_{ad}\delta_{be}\epsilon_{cf}
+3\kappa^{-1}\epsilon_{abc}\epsilon_{def}\delta_{ad}\epsilon_{be}\epsilon_{cf}\nonumber\\
+\epsilon_{abc}\epsilon_{def}\epsilon_{ad}\epsilon_{be}\epsilon_{cf}
=6\kappa^{-3}+3\kappa^{-2}(\epsilon^{abc}\epsilon_{abf})\epsilon_{cf}
+3\kappa^{-1}\bigl(\delta^b_{e}\delta^c_{f}-\delta^b_{f}\delta^c_{e}\bigr)\epsilon_{be}\epsilon_{cf}\nonumber\\
+\epsilon_{abc}\epsilon_{def}\epsilon_{ad}\epsilon_{be}\epsilon_{cf}
=6\kappa^{-3}+6\kappa^{-2}\hbox{tr}\epsilon+3\kappa^{-1}Var\epsilon+6\hbox{det}\epsilon
\end{eqnarray}

\noindent
Putting it all together,

\begin{eqnarray}
q_0=\hbox{det}B\bigl(\Lambda\hbox{det}\Psi+Var\Psi\bigr)+G\Omega_0
=\hbox{det}B\bigl(6\kappa\hbox{det}\Psi+Var\Psi\bigr)+G\Omega_0\nonumber\\
=\hbox{det}B\Bigl(\kappa\bigr(-6\kappa^{-3}-6\kappa^{-2}\hbox{tr}\epsilon-3\kappa^{-1}Var\epsilon-6\hbox{det}\epsilon\Bigr)\nonumber\\
+\hbox{det}B\Bigl(6\kappa^{-2}+{4 \over \kappa}\hbox{tr}\epsilon+Var\epsilon\Bigr)+G\Omega_0\nonumber\\
=\hbox{det}B\Bigl(-2\kappa^{-1}\hbox{tr}\epsilon-2Var\epsilon-6\kappa\hbox{det}\epsilon\Bigr)+G\Omega_0.
\end{eqnarray}

\noindent
The semiclassical part of the Hamiltonian constraint, then, is

\begin{eqnarray}
\label{SMICLS}
q_0=G\Omega_0-2\hbox{det}B\Bigl({{\hbox{tr}\epsilon} \over \kappa}+Var\epsilon+3\kappa\hbox{det}\epsilon\Bigr)
=G\Omega_0-\hbox{det}B\Bigl({{12} \over \Lambda}{\hbox{tr}\epsilon}+2Var\epsilon+\Lambda\hbox{det}\epsilon\Bigr)
\end{eqnarray}

\noindent
Note that the mass dimensions in (\ref{SMICLS}) balance, which is a good check on the algebraic steps.  Rearranging (\ref{SMICLS}) we obtain

\begin{eqnarray}
\label{QQQ111}
\hbox{tr}\epsilon+{\Lambda \over 6}Var\epsilon+{{\Lambda^2} \over {12}}\hbox{det}\epsilon
={{G\Lambda} \over {12\vert{B}\vert}}\Omega_0.
\end{eqnarray}

\subsection{Expansion of matter contribution relative to $\Psi_{Kod}$}

We will assume that the scalar potential $V(\phi)$ can be included as a contribution to the cosmological term $\Lambda$.  Since the Klein--Gordon scalar field contribution to the quantized Hamiltonian constraint is quadratic in conjugate momenta we should expect a contribution both to 
$\Omega_0$ and to $\Omega_1$.  Starting with the classical form of this contribution,

\begin{equation}
\label{KG}
H_{KG}={{\pi^2} \over 2}+{1 \over 2}\partial_{i}\phi\partial_{j}\phi\widetilde{\sigma}^i_{a}\widetilde{\sigma}^j_{a}
\end{equation}

\noindent
Recalling the results from \cite{EYO}, defining $\partial_{i}\phi\partial_{j}\phi=T_{ij}$, we have 
for $\Omega_1$,

\begin{eqnarray}
\label{KG2}
\Omega_1=-{i \over {2G}}{{\partial\pi} \over \partial\phi}
+{1 \over 2}T_{ij}B^i_{e}{\partial \over {\partial{A^j_{a}}}}\Psi_{ae}\nonumber\\
=-{i \over {2G}}{{\partial\pi} \over \partial\phi}
-{1 \over 2}\tau_{ie}{\partial \over {\partial{A^j_{a}}}}\bigl({6 \over \Lambda}\delta_{ae}+\epsilon_{ae}\bigr)
=-{i \over {2G}}{{\partial\pi} \over \partial\phi}-{1 \over 2}\tau_{ie}{\partial \over {\partial{A^j_{a}}}}\epsilon_{ae}
\end{eqnarray}

\noindent
which has the interpretation of an inhomogeneous quantum matter effect proportional to $\partial\pi/\partial\phi$, plus a contribution linear in $\epsilon_{ae}$.\par
\indent
For $\Omega_0$, we have

\begin{eqnarray}
\label{KG3}
\Omega_0={{\pi^2} \over 2}+{1 \over 2}T_{ij}\Psi_{ae}\Psi_{af}B^i_{e}B^j_{f}\nonumber\\
={{\pi^2} \over 2}+{1 \over 2}T_{ij}B^i_{e}B^j_{f}\bigl({6 \over \Lambda}\delta_{ae}+\epsilon_{ae}\bigr)\bigl({6 \over \Lambda}\delta_{af}+\epsilon_{af}\bigr)\nonumber\\
={{\pi^2} \over 2}+{{18} \over {\Lambda^2}}(BB)^{ij}T_{ij}+\epsilon_{ae}({6 \over \Lambda}T_{ij}B^i_{a}B^j_{e})+{1 \over 2}T_{ij}B^i_{e}B^j_{f}\epsilon_{ae}\epsilon_{af}\nonumber\\
=T_{00}+{6 \over \Lambda}\tau_{ae}\epsilon_{ae}+{1 \over 2}\delta_{ab}\tau_{ef}\epsilon_{ae}\epsilon_{bf}
\end{eqnarray}

\noindent
which has the interpretation of an inhomogeneous term corresponding to $T_{00}$, the time-time component of the matter energy momentum tensor coupled to the background metric $(BB)^{ij}$, plus a part linear in $\epsilon_{ae}$, and a quadratic correction.\par

\section{Full expansion of the constraints}
\par
\medskip
\indent 
We are now ready to expand the quantized constraints into a nonlinear system of nine equations in nine unknowns.  There wil be six equations, all linear, from the kinematic constraints and three equations $(q_0=q_1=q_2=0)$ from the Hamiltonian constraint.

\subsection{Total contribution to $q_0$}

Rewriting (\ref{QQQ111}) maintaining all indices explicit,

\begin{eqnarray}
\label{SOL}
\delta_{ae}\epsilon_{ae}+{\Lambda \over 6}\bigl(\delta_{ae}\delta_{bf}-\delta_{af}\delta_{be}\bigr)\epsilon_{ae}\epsilon_{bf}
+{{\Lambda^2} \over {72}}\epsilon_{abc}\epsilon_{def}\epsilon_{ad}\epsilon_{be}\epsilon_{cf}\nonumber\\
={{G\Lambda} \over {12}}\widetilde{\tau}_{00}+{G \over 2}\widetilde{\tau}_{ae}\epsilon_{ae}
+{{G\Lambda} \over {24}}\delta_{ab}\widetilde{\tau}_{ef}\epsilon_{ae}\epsilon_{bf}
\end{eqnarray}

\noindent
The idea is then to transfer all terms linear in $\epsilon_{ae}$ to the left hand side of 
(\ref{SOL}), maintaining all inhomogeneous and nonlinear terms on the right hand side.  This yields

\begin{eqnarray}
\label{SOL1}
\eta_{ae}\epsilon_{ae}=
\bigl(\delta_{ae}-{G \over 2}\widetilde{\tau}_{ae}\bigr)\epsilon_{ae}\nonumber\\
={{G\Lambda} \over {12}}\widetilde{\tau}_{00}-{\Lambda \over 6}\bigl(\delta_{ae}\delta_{bf}-\delta_{af}\delta_{be}
-{G \over 4}\delta_{ab}\widetilde{\tau}_{ef}\bigr)\epsilon_{ae}\epsilon_{bf}
-{{\Lambda^2} \over {72}}\epsilon_{abc}\epsilon_{def}\epsilon_{ad}\epsilon_{be}\epsilon_{cf}\nonumber\\
\end{eqnarray}

\noindent
In (\ref{SOL1}) $\eta^{ae}$ is not to be confused with the spacetime Minkowski metric $\eta^{\mu\nu}$, since the $ae$ indices take their value in $SU(2)_{-}$.  Rather, it should be taken 
as a kind of $SU(2)_{-}$ metric which corresponds to DeSitter spacetime, corrected by the 
$SU(2)_{-}$-projected energy momentum tensor of matter.\par

\subsection{Total contribution to $q_1$}

\noindent
(II) Moving on to the $q_1$ term and dividing through by a factor of $2$,

\begin{eqnarray}
\label{QUA1}
\bigl((\hbox{det}B)\partial^{ae}+2\delta_{ae}\hbox{tr}C\bigr)\epsilon_{ae}
+{\Lambda \over 8}\bigl((\hbox{det}B)\partial^{ef}_{ab}+4C^{ab}_{ef}\bigr)\epsilon_{ae}\epsilon_{bf}
+{1 \over {4i}}{{\partial\pi} \over {\partial\phi}}
+{G \over 4}\tau_{ie}{\partial \over {\partial{A}^a_{i}}}\epsilon_{ae}=0
\end{eqnarray}

\noindent
We now transfer all terms linear in $\epsilon_{ae}$ to the left hand side of (\ref{QUA1}) to obtain

\begin{eqnarray}
\label{QUA2}
\nabla^{bf}\epsilon_{bf}=
\Bigl[\vert{B}\vert\bigl(\delta_{bf}(B^{-1})^a_{i}-(B^{-1})^b_{i}\delta_{af}
+{G \over 4}\delta_{ab}\tau_{if}\bigr){\partial \over {\partial{A}^a_{i}}}+2\delta_{bf}\hbox{tr}C\Bigr]\epsilon_{bf}
={i \over 4}{{\partial\pi} \over {\partial\phi}}
-{\Lambda \over 8}\nabla^{ef}_{ab}\epsilon_{ae}\epsilon_{bf}
\end{eqnarray}

\noindent
where in (\ref{QUA2}) we have made the definitions

\begin{equation}
\label{DEFIN2}
\nabla^{ef}_{ab}=\partial^{ef}_{ab}+4C^{ab}_{ef}
\end{equation}

\noindent
and also

\begin{equation}
C^{ab}_{ef}=\delta^a_{e}C^b_{f}-\delta^b_{e}C^a_{f}.
\end{equation}

\noindent
We have in (\ref{QUA2}) transferred all linear terms to the left hand side, leaving the inhomogeneous (quantum) term and nonlinear terms on the right hand side.  The inhomogeneous quantum matter term $q$ serves as the driving source for functional divergence relative to the pure Kodama state.\par
\noindent
(III) Lastly, the $q_2$ term reads

\begin{equation}
\label{QUA3}
\Delta^{ae}\epsilon_{ae}=0
\end{equation}

\noindent
with $\Delta_{ae}$ defined as in (\ref{CUUUUET}).  Equation (\ref{QUA3}) is a linear operator on the CDJ matrix elements.

\section{Matrix representation of the constraints}
\par
\medskip
\indent
We are now ready to put the constraint equations into matrix form.  First, let us rewrite the equations (\ref{DI}), (\ref{GOOS}), (\ref{SOL1}), (\ref{QUA2}), (\ref{QUA3}) for 
completeness.  They read, leaving the hats off the operators for simplicity,

\begin{eqnarray}
\label{SET}
\epsilon_{ead}\epsilon_{ae}=G\widetilde{\tau}_{0d}\nonumber\\
G^{ae}_{d}\epsilon_{ae}=GQ_d=0\nonumber\\
\eta^{ae}\epsilon_{ae}=G\Bigl({{\Lambda{\widetilde{\tau}}_{00}} \over {12}}\Bigr)-{\Lambda \over 6}\bigl(\delta_{ae}\delta_{bf}-\delta_{af}\delta_{be}-{G \over 4}\delta_{ab}\widetilde{\tau}_{ef}\bigr)\epsilon_{ae}\epsilon_{bf}
-{{\Lambda^2} \over {72}}\epsilon_{abc}\epsilon_{def}\epsilon_{ad}\epsilon_{be}\epsilon_{cf}\nonumber\\
\nabla^{ae}\epsilon_{ae}={i \over 4}{{\partial\pi} \over {\partial\phi}}
-{\Lambda \over 8}\nabla^{eb}_{af}\epsilon_{ae}\epsilon_{bf}\nonumber\\
\Delta^{ae}\epsilon_{ae}=0
\end{eqnarray}

\noindent
the set (\ref{SET}) can be written in the form

\begin{equation}
\label{SOF}
O_{ab}^{cd}\epsilon_{cd}=GQ_{ab}+\xi_{ab}.
\end{equation}

\noindent
where $G$ is Newton's gravitational constant.\par
\indent  
In (\ref{SOF}) $O$ is a linear transformation acting on the CDJ deviation matrix, which can be represented by a nine by nine matrix of differential operators.  These differential operators can be thought of as vector fields with respect to directions in the base space $B_{GKod}$ of the fibre bundle $E_{GKod}$.  In a sense, the operator $O$ acts as the analogue of a structure group for the fibre bundle, which transforms the components of the fibre (section) $\epsilon_{ab}$ in 
a manner prescribed by the dynamics of quantum gravity as encoded in the quantized constraints.  Schematically,

\begin{equation}
\label{FIB}
E_{GKod}\equiv\bigl(\Sigma,\Gamma,\epsilon_{ae}(\Sigma,\Gamma)\bigr)
\longrightarrow\bigl(\Sigma,\Gamma,\epsilon_{ab}(\Sigma,\Gamma),O_{ab}^{cd}(\Gamma;\partial_{\Sigma},\partial_A)\bigr)
\end{equation}

\noindent
where we have adopted the notation

\begin{equation}
\label{FIB1}
O_{ab}^{cd}(\Gamma;\partial_{\Sigma},\partial_A)\equiv{O}_{ab}^{cd}\Bigl(\phi^{\alpha},A^a_i;{\partial \over {\partial{x^i}}},{\partial \over {\partial{A^a_i}}}\Bigr)
\end{equation}

\noindent
for $x^i=(x,y,z)\rightarrow(t^1,t^2,t^3)$ in $\Sigma$ and $(A^a_i,\phi^{\alpha})$ in $\Gamma$.  In (\ref{FIB1}) we have highlighted the model-specific matter dependence of the operator $O$.  The matrix representations in (\ref{SOF}) are given, first for the `structure' transformation matrix $O$, by

\begin{displaymath}
O_{ab}^{cd}=
\left(\begin{array}{ccccccccc}
-1 & 0 & 0 & 1 & 0 & 0 & 0 & 0 & 0\\
0 & -1 & 0 & 0 & 1 & 0 & 0 & 0 & 0\\
0 & 0 & -1 & 0 & 0 & 1 & 0 & 0 & 0\\
(\hat{G}_1)^{21} & (\hat{G}_1)^{32} & (\hat{G}_1)^{13} & (\hat{G}_1)^{12} & (\hat{G}_1)^{23} & (\hat{G}_1)^{31} & (\hat{G}_1)^{11} & (\hat{G}_1)^{22} & (\hat{G}_1)^{33}\\ 
(\hat{G}_2)^{21} & (\hat{G}_2)^{32} & (\hat{G}_2)^{13} & (\hat{G}_2)^{12} & (\hat{G}_2)^{23} & (\hat{G}_2)^{31} & (\hat{G}_2)^{11} & (\hat{G}_2)^{22} & (\hat{G}_2)^{33}\\ 
(\hat{G}_3)^{21} & (\hat{G}_3)^{32} & (\hat{G}_3)^{13} & (\hat{G}_3)^{12} & (\hat{G}_3)^{23} & (\hat{G}_3)^{31} & (\hat{G}_3)^{11} & (\hat{G}_3)^{22} & (\hat{G}_3)^{33}\\ 
\eta^{21} & \eta^{32} & \eta^{13} & \eta^{12} & \eta^{23} & \eta^{31} & \eta^{11} & \eta^{22} & \eta^{33}\\
\hat{\nabla}^{21} & \hat{\nabla}^{32} & \hat{\nabla}^{13} & \hat{\nabla}^{12} & \hat{\nabla}^{23} & \hat{\nabla}^{31} 
& \hat{\nabla}^{11} & \hat{\nabla}^{22} & \hat{\nabla}^{33}\\ 
\hat{\Delta}^{21} & \hat{\Delta}^{32} & \hat{\Delta}^{13} & \hat{\Delta}^{12} & \hat{\Delta}^{23} & \hat{\Delta}^{31} 
& \hat{\Delta}^{11} & \hat{\Delta}^{22} & \hat{\Delta}^{33}\\ 
\end{array}\right).
\end{displaymath}

\noindent
It will be important to keep note of the mass dimensions of the components of the operator $O$ as displayed above.  This will be of significance when we invert the operators.  The
mass dimensions are given by

\begin{equation}
\label{DIM}
[1]=0;~~[G_{a}^{bc}]=3;~~[\eta^{ab}]=0;~~[\nabla^{ab}]=3;~~[\Delta^{ab}]=0;~~[\epsilon_{ae}]=-2
\end{equation}

\noindent
The mass dimensions must balance on both sides of (\ref{SOF}).\par
\indent  
The CDJ deviation matrix $\epsilon_{ab}$ is given by

\begin{displaymath}
\epsilon_{ab}=
\left(\begin{array}{c}
\epsilon_{21}\\
\epsilon_{32}\\
\epsilon_{13}\\
\epsilon_{12}\\
\epsilon_{23}\\
\epsilon_{31}\\
\epsilon_{11}\\
\epsilon_{22}\\
\epsilon_{33}\\
\end{array}\right)
\end{displaymath}

\noindent
Note how the matrix elements $\epsilon_{ae}$ are arranged such that the correct linear transformation $O$ is implemented.  The inhomogeneous term $Q_{ab}$ is given by

\begin{displaymath}
Q_{ab}=
\left(\begin{array}{c}
Q_{21}\\
Q_{32}\\
Q_{13}\\
Q_{12}\\
Q_{23}\\
Q_{31}\\
Q_{11}\\
Q_{22}\\
Q_{33}\\
\end{array}\right)
=
\left(\begin{array}{c}
\widetilde{\tau}_{01}\\
\widetilde{\tau}_{02}\\
\widetilde{\tau}_{03}\\
Q_1\\
Q_2\\
Q_3\\
(\Lambda/12)\widetilde{\tau}_{00}\\
q\\
0\\
\end{array}\right)
\end{displaymath}

\noindent
The quantities $\tau_{0a}$ are the $SU(2)_{-}$ analogue of the time-space part of the matter energy momentum tensor in dimensionless units, given by

\begin{equation}
\label{SCALE1}
\widetilde{\tau}_{0a}
={{B^i_{a}H_i} \over {\hbox{det}B}}={{B^i_{a}} \over {\hbox{det}B}}T_{0i}
\end{equation}

\noindent
The quantity $Q_a$ is the $SU(2)_{-}$ charge, and is of mass dimension $[Q_a]=3$.  It is zero for the Klein--Gordon field, but we had put it in for explicative purposes.  The quantity $q$ is given by

\begin{equation}
\label{QUEUE}
q={i \over {4G}}{{\partial\pi} \over {\partial\phi}}
\end{equation}

\noindent
and has mass dimension $[q]=3$.  This is a quantum effect which manifests itself already at the linearized level of the constraints.\par
\indent
All sources associated with the Einstein's classical metric relativity, namely the dynamic components of the stress-energy tensor $T_{0\mu}$, have been made dimensionless.  However, the sources indigenous to classical Ashtekar variables and quantized matter effects, $Q_a$ and $q$ are not.  This imbalance needs to be rectified, since the elements of $O_{ab}^{cd}$ mix the components of 
$Q_{ab}$ together in (\ref{SOF}).  It will be rectified during the process of inversion of the matrix $\hat{O}$.\par
\indent
Our method of solution to (\ref{SOF}) will be by iteration.  First we solve the linear inhomogeneous equation with the error $\xi_{ab}$ absent and then we compute the error.  This
error contributes to the inhomogeneous source term $Q_{ab}$ for the next stage of the iteration, etc.  In the process we should build a power series in the source $Q_{ab}$ and
then establish criteria for convergence of the series.  The inversion of the matrix $\hat{O}$ appears problematic since its elements consist of noncommuting differential operators.  So one requirement for the existence of solutions to the quantum constraints is that the inverse be well-defined.\par

\subsection{The error vector}

Before we attempt inversion of $O_{ab}^{cd}$, let us write down the error vector $\xi_{ab}$.  The error vector encodes the nonlinearities inherent in the quantized Hamiltonian
constraint and forms the driving mechanism by which the CDJ deviation tensor should converge, under the flow induced by the linear operator $O$, to a unique solution.  A lot about
the error vector, which is model specific, hopefully may be used to assess quantities such as fixed points and stability or instability of the solution.  The orbit or flow of 
$\epsilon_{ae}$ within the fibre of the fibre bundle $E_{GKod}$ can be interpreted as a kind of renormalization of the $SU(2)$ metric $\Psi_{ab}$ relative to DeSitter 
spacetime.  The error vector is given by

\begin{displaymath}
\xi_{ab}=
\left(\begin{array}{c}
\xi_{21}\\
\xi_{32}\\
\xi_{13}\\
\xi_{12}\\
\xi_{23}\\
\xi_{31}\\
\xi_{11}\\
\xi_{22}\\
\xi_{33}\\
\end{array}\right)
=
\left(\begin{array}{c}
0\\
0\\
0\\
0\\
0\\
0\\
\xi_1\\
\xi_2\\
0\\
\end{array}\right)
\end{displaymath}

\noindent
In tensor notation this is given by

\begin{equation}
\label{ERROR}
\xi_{ab}=\delta_{a1}\delta_{b1}\xi_1[\epsilon_{ae},\phi^{\alpha}]+\delta_{a2}\delta_{b2}\xi_2[\epsilon_{ae},\phi^{\alpha}]
\end{equation}

\noindent
where $\xi_1$ and $\xi_2$, are given by

\begin{eqnarray}
\label{ERROR1}
\xi_{1}=-\Bigl(\Lambda{I}^{aebf}\epsilon_{ae}\epsilon_{bf}+\Lambda^{2}E^{abcdef}\epsilon_{ad}\epsilon_{be}\epsilon_{cf}\Bigr);~~
\xi_2=-{\Lambda \over 8}\nabla^{eb}_{af}\epsilon_{ae}\epsilon_{bf}
\end{eqnarray}

\noindent
where we have defined

\begin{eqnarray}
\label{DEFN1}
{I}^{aebf}={1 \over 6}\bigl(\delta_{ae}\delta_{bf}-\delta_{af}\delta_{be}
-{G \over 4}\delta_{ab}\widetilde{\tau}_{ef}\bigr);~~E^{abcdef}={1 \over 72}\epsilon^{abc}\epsilon^{def}
\end{eqnarray}

\noindent
The mass dimensions are given by 

\begin{equation}
[\xi_1]=-2;~~[\xi_2]=1
\end{equation}

\noindent
Note that $\xi_2$ is independent of the model and $\xi_1$ is model specific, scaled relative to its model-free
counterpart by terms of order $\widetilde{\tau}_{ae}$.

\section{Inversion of $O_{ab}^{cd}$}
\par
\medskip
\indent
One strategy to invert the nine by nine matrix $O_{ab}^{cd}$ is to split $O$ into diagonal and off-diagonal parts $\hat{O}=\hat{D}+\hat{R}$, with $\hat{D}$ diagonal and the remainder 
$\hat{R}$ off-diagonal.  We will assume that the diagonal part is invertible since there is no commutation of differential operators required.  Since the matrix $\hat{O}$ consists of differential operators, either with respect to $\Sigma$ or $\Gamma$, its inverse will consist of path-ordered integration along directions in the base space $B_{GKod}=(\Sigma,\Gamma)$ of the fibre bundle.  Inversion of the diagonal part is tantamount to finding a basis with respect to which the fibre bundle structure $\bigl(\Sigma,\Gamma,\Psi_{ab}(\Sigma,\Gamma)\bigr)$ is manifestly trivialized.  That this is possible is a direct consequence of the commutativity of the Gauss' law constraint with the quantized Hamiltonian constraint \cite{ASH},\cite{ASH1},

\begin{equation}
\bigl[\hat{G}_a,\hat{H}\bigr]=0.
\end{equation}

\indent  
We then compose the inverted diagonal part with the nondiagonal part to form a composite operator $\hat{U}$, which acts on the right hand side.  We express the result formally as a 
geometric series of operators $\hat{U}=\hat{D}^{-1}\hat{R}$, which can be put in path-ordered or functional-ordered form in analogy to the results of \cite{EYO} regarding the Gauss' law constraint. Schematically, we manipulate (\ref{SOF}) into the form,

\begin{equation}
\label{SOLV}
O\epsilon=GQ+\xi(\epsilon).
\end{equation}

\noindent
First we solve the linear inhomogeneous part of (\ref{SOLV})

\begin{equation}
\label{SOLV1}
O\epsilon\sim{S}\longrightarrow\epsilon\sim{O}^{-1}S.
\end{equation}

\noindent
where $S=GQ$ represents the source vector consisting of matter-dependent terms.  We do this schematically via the steps, omitting the hats for simplicity,

\begin{eqnarray}
\label{SOLV2}
\zeta=(D+R)\epsilon=S\longrightarrow{D}\epsilon=-R\epsilon+S\nonumber\\
\epsilon=-(D^{-1}R)\epsilon+D^{-1}S\longrightarrow(1+U)\epsilon=D^{-1}S=Q^{\prime}\nonumber\\
\epsilon\equiv(1+U)^{-1}Q^{\prime}=\Bigl(\sum_{n=1}^{\infty}(-1)^{n}U^n\Bigr)Q^{\prime}\equiv\bigl[\hat{P}e^{-U}\bigr]Q^{\prime}
\end{eqnarray}

where the diagonal part is given by

\begin{displaymath}
D_{ab}^{cd}=
\left(\begin{array}{ccccccccc}
1 & 0 & 0 & 0 & 0 & 0 & 0 & 0 & 0\\
0 & 1 & 0 & 0 & 0 & 0 & 0 & 0 & 0\\
0 & 0 & 1 & 0 & 0 & 0 & 0 & 0 & 0\\
0 & 0 & 0 & (\hat{G}_1)^{12} & 0 & 0 & 0 & 0 & 0\\ 
0 & 0 & 0 & 0 & (\hat{G}_2)^{23} & 0 & 0 & 0 & 0\\ 
0 & 0 & 0 & 0 & 0 & (\hat{G}_3)^{31} & 0 & 0 & 0\\ 
0 & 0 & 0 & 0 & 0 & 0 & 1 & 0 & 0\\
0 & 0 & 0 & 0 & 0 & 0 & 0 & \hat{\nabla}^{22} & 0\\ 
0 & 0 & 0 & 0 & 0 & 0 & 0 & 0 & \hat{\Delta}^{33}\\ 
\end{array}\right)
\end{displaymath}

\noindent
and the off-diagonal remainder $R$ is given by

\begin{displaymath}
R_{ab}^{cd}=
\left(\begin{array}{ccccccccc}
0 & 0 & 0 & 1 & 0 & 0 & 0 & 0 & 0\\
0 & 0 & 0 & 0 & 1 & 0 & 0 & 0 & 0\\
0 & 0 & 0 & 0 & 0 & 1 & 0 & 0 & 0\\
(\hat{G}_1)^{21} & (\hat{G}_1)^{32} & (\hat{G}_1)^{13} & 0 & (\hat{G}_1)^{23} & (\hat{G}_1)^{31} & (\hat{G}_1)^{11} & (\hat{G}_1)^{22} & (\hat{G}_1)^{33}\\ 
(\hat{G}_2)^{21} & (\hat{G}_2)^{32} & (\hat{G}_2)^{13} & (\hat{G}_2)^{12} & 0 & (\hat{G}_2)^{31} & (\hat{G}_2)^{11} & (\hat{G}_2)^{22} & (\hat{G}_2)^{33}\\ 
(\hat{G}_3)^{21} & (\hat{G}_3)^{32} & (\hat{G}_3)^{13} & (\hat{G}_3)^{12} & (\hat{G}_3)^{23} & 0 & (\hat{G}_3)^{11} & (\hat{G}_3)^{22} & (\hat{G}_3)^{33}\\ 
\eta^{21} & \eta^{32} & \eta^{13} & \eta^{12} & \eta^{23} & \eta^{31} & 0 & \eta^{22} & \eta^{33}\\
\hat{\nabla}^{21} & \hat{\nabla}^{32} & \hat{\nabla}^{13} & \hat{\nabla}^{12} & \hat{\nabla}^{23} & \hat{\nabla}^{31} 
& \hat{\nabla}^{11} & 0 & \hat{\nabla}^{33}\\ 
\hat{\Delta}^{21} & \hat{\Delta}^{32} & \hat{\Delta}^{13} & \hat{\Delta}^{12} & \hat{\Delta}^{23} & \hat{\Delta}^{31} 
& \hat{\Delta}^{11} & \hat{\Delta}^{22} & 0\\ 
\end{array}\right).
\end{displaymath}

In order to solve the linear part of the constraint equations it is necessary only to invert five differential operators, 

\begin{eqnarray}
\label{DIA}
D^{-1}=Diag\bigl(-1,-1,-1,(\hat{G}_1)^{12},(\hat{G}_2)^{23},(\hat{G}_3)^{31},\eta^{11},\hat{\nabla}_{22},\hat{\Delta}_{33}\bigr)^{-1}\nonumber\\
=Diag\bigl(-1,-1,-1,(\hat{G}_1^{12})^{-1},(\hat{G}_2^{23})^{-1},(\hat{G}_3^{31})^{-1},{1 \over {\eta^{11}}},(\hat{\nabla}_{22})^{-1},(\hat{\Delta}_{33})^{-1}\bigr)
\end{eqnarray}

\noindent
where $\eta^{11}$ is a c-number.  The five differential operators correspond to five linearly independent directions within the base space $B_{GKod}\equiv(\Sigma,\Gamma)$, namely three directions in 3-space $\Sigma$ (due to Gauss' law) and two directions in functional space $\Gamma$ 
(due to $q_1$ and $q_2$).  These directions form a five-dimensional subspace of $B_{GKod}$.  Any boundary conditions due to integration in these directions should be included as part of the inversion process (\ref{DIA}), but we will in practice take them to be zero, since we are comparing $\Psi_{GKod}$ to $\Psi_{Kod}$, the latter for which $\epsilon_{ae}=0~\forall~(x,{A^a_i}(x))$.\par
\indent
The inversion of $\hat{D}$ implements two main effects:\par
\noindent
(i) It transforms $Q_{ab}$, which contains some dimensionful components associated with Gauss' law charges and quantum terms of the Hamiltonian constraint, into 
$Q^{\prime}_{ab}$ which is dimensionless.  This is given by $Q^{\prime}=D^{-1}Q$

\begin{displaymath}
Q^{\prime}_{ab}=
\left(\begin{array}{c}
Q^{\prime}_{21}\\
Q^{\prime}_{32}\\
Q^{\prime}_{13}\\
Q^{\prime}_{12}\\
Q^{\prime}_{23}\\
Q^{\prime}_{31}\\
Q^{\prime}_{11}\\
Q^{\prime}_{22}\\
Q^{\prime}_{33}\\
\end{array}\right)=
\left(\begin{array}{c}
\widetilde{\tau}_{01}\\
\widetilde{\tau}_{02}\\
\widetilde{\tau}_{03}\\
(\hat{G}_1^{12})^{-1}Q_1\\
(\hat{G}_2^{23})^{-1}Q_2\\
(\hat{G}_3^{31})^{-1}Q_3\\
\tau_{00}/\eta^{11}\\
(\hat{\nabla}_{22})^{-1}q\\
(\hat{\Delta}_{33})^{-1}0\\
\end{array}\right)
\end{displaymath}

\noindent
At this stage all elements of the source term of (\ref{SET}), being dimensionless, have been placed on equal footing.  A convenient physical interpretation is that $q$ and $Q_a$, external structures with respect to Einstein's classical metric gravity are initially unsynchronized with respect to 
$T_{0\mu}$ in $B_{GKod}$.  Nonlocal transformations of $Q_a$ and $q$, respectively with respect to $\Sigma$ and $\Gamma$, are required to bring them into synchronicity.  In this sense the energy $\tau_{00}$ has been brought into synchronicity as well, albeit through a local transformation.\par
\noindent
(ii) The inversion of $\hat{D}$ also determines $\hat{U}$, which can be seen as a normalized version of $\hat{D}$, the rows normalized by their respective diagonal elements. 
$\hat{U}\equiv{U}_{ab}^{cd}$ acts on all components of the `normalized' source $Q^{\prime}_{ab}$ identically.  In effect, Einstein's metric relativity, Ashtekar variables and 
quantum theory have all been placed on the same footing.  The matrix $U$ is given by

\begin{displaymath}
U_{ab}^{cd}=
\left(\begin{array}{c}
-\delta_{4c}\delta_{4d}\\
-\delta_{5c}\delta_{5d}\\
-\delta_{6c}\delta_{6d}\\
(\hat{G}_1^{12})^{-1}\hat{G}_1^{cd}-\delta_{c1}\delta_{d2}\\
(\hat{G}_2^{23})^{-1}\hat{G}_2^{cd}-\delta_{c2}\delta_{d3}\\
(\hat{G}_3^{31})^{-1}\hat{G}_3^{cd}-\delta_{c3}\delta_{d1}\\
\eta^{cd}/\eta^{11}-\delta_{c1}\delta_{d1}\\
(\hat{\nabla}_{22})^{-1}\hat\nabla^{cd}-\delta_{c2}\delta_{d2}\\
(\hat{\Delta}_{33})^{-1}\hat\Delta^{cd}-\delta_{c3}\delta_{d3}\\
\end{array}\right)
\end{displaymath}

\noindent
The mass dimensions at the linearized level of the flow equation 

\begin{equation}
\label{FLOW}
\epsilon_{ab}=GU_{ab}^{cd}\bigl(Q^{\prime}_{cd}+\xi_{ab}\bigr)
\end{equation}

\noindent
are given by

\begin{equation}
[\epsilon_{ae}]=-2;~~[U_{ab}^{cd}]=0;~~[Q^{\prime}_{ab}]=0.
\end{equation}

\subsection{Path-ordered integration in the unified base space}

Before we attempt a general solution of the flow equation let us note a general property of the solution.  The inversion of differential operators with respect to the five orthogonal directions in $B_{GKod}$ corresponds to path-ordered integration with respect to these directions, which traces out a path in a five-dimensional subspace.\par
\indent  
Let us unify the base space of the bundle into the generalized form $(\Sigma,\Gamma)\rightarrow\chi$, where $\chi=\chi_{ab}$ is the variable of integration in question, whether it be one of the three spatial coordinates $x^i$ or one of the two components of the connection 
$A^a_i$.  Then associated with each $\chi_{ab}$ is a Green's function with respect to propagation in position space for each function for all functions, or propagation in functional space for a fixed position for each position.  Let $\nu_{ab}^{cd}$ be the set of variables of differentiation comprising the matrix operator $\hat{R}$, one variable of differentiation for each nontrivial element of R, each forming a vector field in $B_{GKod}$.  The net effect of $\hat{R}$ upon the source $Q^{\prime}$ is to mix their differentiated components together in a specific way.  There are a total of $5\times{9}-5=40$ such variables $\nu$, but only 
$3+3\times{3}$ independent variables per spatial point in the full theory (three from space 
$\Sigma$ and nine from $A^a_i$).  Therefore the vector fields form an overcomplete basis.\par
\indent
The general solution to the inversion of $\hat{D}$ then is given, assuming trivial boundary conditions (which would correspond to the pure Kodama state, with a fixed point $\epsilon_{ae}=0$), by

\begin{equation}
\label{PROP}
(\hat{D}^{-1})_{ab}^{cd}S_{cd}[\chi]
=\int^{\chi_{ab}}_{\beta_{ab}}d\chi^{\prime}_{ab}\hat{K}_{ab}\bigl(\chi_{ab},\chi^{\prime}_{ab}\bigr)S_{cd}(\chi^{\prime}_{ab}),
\end{equation}

\noindent
where $\beta_{ab}=\beta_{ab}(x)$ is a function for which the `functional' boundary condition on 
$\epsilon_{ae}$ is determined.  This is analogous to choosing the boundary, in usual differential equations, upon which the boundary conditions on the function being solved for are specified in order to produce a unique solution.  In the field-theoretical analogue the choice can be made judiciously, for example such as to impose certain desired conditions on 
$\Psi_{GKod}$.\par
\indent
The two-index object $\hat{K}_{ab}$ in (\ref{PROP}) is the propagator or `generalized' Green's function or functional heat kernel associated with inversion of the component of $\hat{D}$ in the direction of $\chi_{ab}$.  Incidentally, the action of $\hat{U}=\hat{D}^{-1}\hat{R}$ on a source vector can be interpreted schematically in the following way, if we suppress all dependence upon any spectator variables $\alpha$ in $B_{GKod}$.  $\hat{R}$ corresponds to a generator of transformations of the source term $S\equiv{S}[\chi,\nu]$ in the $\nu$ directions and 
$\hat{D}^{-1}$ corresponds to the inverse transformation in the $\chi$ directions which are in general different.  The effect of the noncommutativity of these transformations is to change the $\nu$ dependence by a $\chi$-dependent amount, or to mix the coordinates.  The path-ordered exponentiation of this transformation can be seen as a kind of `generalized' translation of the source within $B_{Kod}$.  Schematically, 

\begin{equation}
\label{PROPA}
\bigl[\hat{P}e^{-U}\bigr]S[\chi,\nu]\sim{S}[\chi,\nu+\eta(\chi,\nu;\alpha)]
\end{equation}

\noindent
for some nonlocal function $\eta$.  Hence the effect of (\ref{PROPA}) is to drag the source $S$ from one position in $B_{GKod}$ to another with each iteration, tracing out an orbit in $B_{GKod}$.  The unexponetiated action is given, written in components, by

\begin{eqnarray}
\label{PROP1}
U_{ab}^{cd}S_{cd}=(D^{-1})_{ab}^{cd}R_{cd}^{ef}S_{ef}\nonumber\\
\equiv
\int^{\chi_{ab}}_{\beta_{ab}}d\chi^{\prime}_{ab}\hat{K}_{ab}\bigl(\chi_{ab},\chi^{\prime}_{ab}\bigr)\hat{R}^{ef}_{ab}[\nu_{ef}^{cd}]S_{cd}(\chi^{\prime}_{ab},\nu_{fe}^{cd})\nonumber\\
\end{eqnarray}

\indent
We will now write down a general solution for the constraints of general relativity.  To recapitulate, the constraints can be written relative to a nine-dimensional vector space 
in some irreducible basis.  The fibre of $E_{GKod}$, namely the CDJ deviation matrix 
$\epsilon_{ae}$ can be written as the components of a 9-vector in this space

\begin{equation}
\label{FIBRE}
E\equiv\bigl(\Sigma,\Gamma(\Sigma),\epsilon_{ab}(\Gamma),\Sigma)\bigr)
\longrightarrow\bigl(x,A^a_{i}(x),\phi^{\alpha}(x),\epsilon_{ab}(x,A^a_{i},\phi^{\alpha})\bigr).
\end{equation}

\section{Iterative general solution for $\epsilon_{ae}$}
\par
\medskip

\noindent
The full set of quantized constraints can be written in the form

\begin{equation}
\label{UNRED}
O_{ab}^{cd}\epsilon_{cd}=GQ^{\prime}_{ab}+\xi_{ab}
\end{equation}

The correction terms $\xi_{ab}$ are suppressed relative to the linear parts by small numerical factors.  Moving on to the construction of the solution, it will be convenient first to express the error vector in tensor notation

\begin{equation}
\xi_{ab}=\delta_{a1}\delta_{b1}\xi_{11}+\delta_{a2}\delta_{b2}\xi_{22}+\delta_{a3}\delta_{b3}\xi_{33}
\end{equation}

The straightforward method for solving the constraint equations is to solve the linear part first.  To linear order one has

\begin{equation}
\label{UNRED1}
O_{ab}^{cd}(\epsilon_{cd})_{(0)}=G(Q^{\prime}_{ab})_{(0)}\longrightarrow(\epsilon_{ab})_{(0)}=GU_{ab}^{cd}(Q^{\prime}_{cd})_{(0)}.
\end{equation}

\noindent
Note in (\ref{UNRED1}) that we have taken initial error $(\xi_{ab})_{(0)}=0$.  We also have taken this invertibility as given, since we have from the previous section an explicit expansion of $U$ to all orders, as well as its interpretation as a kind of nonlinear group transformation.  The CDJ matrix to this order is given by

\begin{equation}
\label{CDJ1}
\Psi_{ab}=-{6 \over \Lambda}\Bigl(\delta_{ab}+{1 \over 6}(G\Lambda)\hat{U}_{ab}^{cd}Q^{\prime}_{cd}\Bigr)
\end{equation}    

\noindent
Note that all terms in brackets in (\ref{CDJ1}) are dimensionless.  The coupling constant 
$G\Lambda$ is also dimensionless in units for which $\hbar=1$, though it does not contain a factor of $\hbar$.\par
\indent
Next we compute the error tensor, using the solution for $\epsilon$ from (\ref{UNRED1}), along with the matter input $\Omega$ to yield 
$(\xi_{ab})_{(0)}=\xi_{ab}\bigl[GU^{-1}Q\bigr]$.  Note that this expression also takes into consideration the explicit dependence of $\Omega$ upon $\epsilon_{ae}$.  This is given,
using (\ref{ERROR}), (\ref{ERROR1}), and (\ref{DEFN1}) by

\begin{eqnarray}
\label{ERROR3}
(\xi_{gh})_{(1)}=-\delta_{g1}\delta_{h1}
\Bigl(\Lambda{G}^{2}{I}^{aebf}\hat{U}_{ae}^{a^{\prime}e^{\prime}}\hat{U}_{bf}^{b^{\prime}f^{\prime}}Q^{\prime}_{a^{\prime}e^{\prime}}Q^{\prime}_{b^{\prime}f^{\prime}}\nonumber\\
+\Lambda^{2}G^{3}E^{abcdef}\hat{U}_{ad}^{a^{\prime}d^{\prime}}\hat{U}_{be}^{b^{\prime}e^{\prime}}\hat{U}_{cf}^{c^{\prime}f^{\prime}}
Q^{\prime}_{a^{\prime}d^{\prime}}Q^{\prime}_{b^{\prime}e^{\prime}}Q^{\prime}_{c^{\prime}f^{\prime}}\Bigr)\nonumber\\
-{\Lambda \over 8}G^{2}\delta_{g2}\delta_{h2}\Bigl(\nabla^{eb}_{af}(\hat{U}_{ae}^{a^{\prime}e^{\prime}}\hat{U}_{bf}^{b^{\prime}f^{\prime}})
+\hat{U}_{ae}^{a^{\prime}e^{\prime}}\hat{U}_{bf}^{b^{\prime}f^{\prime}}\nabla^{eb}_{af}\Bigr)
Q^{\prime}_{a^{\prime}e^{\prime}}Q^{\prime}_{b^{\prime}f^{\prime}}\nonumber\\
={1 \over \Lambda}\Bigl[(G\Lambda)^{2}\hat{U}_{gh}^{a_{1}b_{1}a_{2}b_{2}}Q^{\prime}_{a_{1}b_{1}}Q^{\prime}_{a_{2}b_{2}}
+(G\Lambda)^{3}\hat{U}_{gh}^{a_{1}b_{1}a_{2}b_{2}a_{3}b_{3}}Q^{\prime}_{a_{1}b_{1}}Q^{\prime}_{a_{2}b_{2}}Q^{\prime}_{a_{3}b_{3}}\Bigr]
\end{eqnarray}

\noindent
where in (\ref{ERROR3}) we have made the identifications

\begin{eqnarray}
\label{ERROR4}
\hat{U}_{gh}^{a_{1}b_{1}a_{2}b_{2}}=-\delta_{g1}\delta_{h1}{I}^{aebf}\hat{U}_{ae}^{a_{1}b_{1}}\hat{U}_{bf}^{a_{2}b_{2}}\nonumber\\
-\delta_{g2}\delta_{h2}\Bigl(\hat{\nabla}^{eb}_{af}(\hat{U}_{ae}^{a_{1}b_{1}}\hat{U}_{bf}^{a_{2}b_{2}})
+\hat{U}_{ae}^{a_{1}b_{1}}\hat{U}_{bf}^{a_{2}b_{2}}\hat{\nabla}^{eb}_{af}\Bigr),
\end{eqnarray}

\noindent
which is an operator that acts in the appropriate fashion on the $Q^{\prime}$ vectors, and likewise

\begin{eqnarray}
\label{ERROR5}
\hat{U}_{gh}^{a_{1}b_{1}a_{2}b_{2}a_{3}b_{3}}
=\delta_{g2}\delta_{h2}E^{abcdef}\hat{U}_{ad}^{a_{1}b_{1}}\hat{U}_{be}^{a_{2}b_{2}}\hat{U}_{cf}^{a_{3}b_{3}}
\end{eqnarray}

\noindent
Also, we have omitted the subscripts on $(Q_{ab})_{(0)}$ in (\ref{ERROR3}) and (\ref{ERROR4}) for simplicity.  Taking the error vector (\ref{ERROR3}) for this stage, which is 
now explicitly a function of the matter source variables, write the new constraint as

\begin{eqnarray}
\label{UNRED2}
O_{ab}^{cd}(\epsilon_{cd})_{(1)}=G(Q^{\prime}_{ab})_{(0)}+(\xi_{ab})_{(1)}\bigl[GUQ^{\prime}_{(0)}\bigr]\nonumber\\
\longrightarrow(\epsilon_{ab})_{(1)}=GU_{ab}^{cd}(Q^{\prime}_{cd})_{(0)}=GU_{ab}^{cd}Q^{\prime}_{cd}+U_{ab}^{cd}(\xi_{cd})_{(1)}\bigl[GUQ^{\prime}_{(0)}\bigr]
\end{eqnarray}

\noindent
whereupon a solution is obtained for the second stage of iteration.  The corresponding CDJ matrix is given by

\begin{eqnarray}
\label{CDJ2}
\Psi_{ab}=-{6 \over \Lambda}\Bigl(\delta_{ab}+{1 \over 6}(G\Lambda)\hat{U}_{ab}^{cd}Q^{\prime}_{cd}
+{1 \over 6}(G\Lambda)^{2}\hat{U}^{gh}_{ab}\hat{U}_{gh}^{a_{1}b_{1}a_{2}b_{2}}Q^{\prime}_{a_{1}b_{1}}Q^{\prime}_{a_{2}b_{2}}\nonumber\\
+{1 \over 6}(G\Lambda)^{3}\hat{U}^{gh}_{ab}\hat{U}_{gh}^{a_{1}b_{1}a_{2}b_{2}a_{3}b_{3}}Q^{\prime}_{a_{1}b_{1}}Q^{\prime}_{a_{2}b_{2}}Q^{\prime}_{a_{3}b_{3}}\Bigr)
\end{eqnarray}

The process is then repeated via the identification

\begin{equation}
\label{UNRED3}
(\epsilon_{ab})_{(n+1)}=GU_{ab}^{cd}(Q^{\prime}_{cd})_{(n+1)}=(\epsilon_{ab})_{(n+1)}+U_{ab}^{cd}(\xi_{cd})_{(n)}\bigl[(\epsilon_{ab})_{(n)}\bigr]
\end{equation}

\noindent
and the resulting series analyzed either for convergence as well as for stability and nonperturbative behavior relative to the fixed point $\Psi_{Kod}$ for which $(\epsilon_{ab})_{n}=0~\forall{n}$.  At each stage in the iteration procedure (\ref{UNRED3}) one is solving a linear equation which can be expressed explicilty in terms of $(Q^{\prime}_{ab})_{0}$.  The interpretation is that all the quantities making up the dimensionless vector $Q^{\prime}$ are renormalized.  This includes the time-space part of the energy momentum tensor of matter $T_{0\mu}$ and the $SU(2)_{-}$ charge $Q_a$ as well as $q$.  The space-space part $T_{ij}$ is not renormalized since it is not part of the source, however it does affect the renormalization of the source.\par

\subsection{Sophisticated interpretation}

Now we move on to the sophisticated interpretation.  The idea is to define a new 9-vector $\zeta_{ab}$ forming the left hand side of (\ref{UNRED}).  The interpretation is that if $Q^{\prime}_{ab}$ represents the unrenormalized source, then $\zeta_{ab}$ is the fully renormalized counterpart, given by

\begin{equation}
\label{UNREDUCED}
\zeta_{ab}=O_{ab}^{cd}\epsilon_{cd}=GQ_{ab}^{\prime}+\xi_{ab}.
\end{equation}

\noindent
One then expresses the right hand side of (\ref{UNREDUCED}) in terms of $\zeta_{ab}$.  A similar iterative procedure is then performed, except this time with a convenient physical interpretation.  First we must calculate the error vector.  We will illustrate this to first-order, for simplicity, in order to illustrate the basic idea.  We will maintain all indices explicit in this simplified example.\par
\indent
The procedure is as follows:\par
\noindent
(i) We must first eliminate $\epsilon_{ae}$ in favor of $\zeta_{ab}$, which requires the inversion

\begin{equation}
\label{UNREDUCED1}
\epsilon_{ab}=U_{ab}^{cd}\zeta_{cd}.
\end{equation}

\noindent
(ii) We must now express the components of the error vector explicitly in terms of 
$\zeta_{ab}$.  Let us redefine them, from (\ref{ERROR1})

\begin{eqnarray}
\label{ERROR11}
\xi_{1}
=-\Lambda{I}^{ae}_{bf}\epsilon_{ae}\epsilon_{bf}
-{{\Lambda^2} \over {72}}\epsilon_{abc}\epsilon_{def}\epsilon_{ad}\epsilon_{be}\epsilon_{cf}
\end{eqnarray}

\noindent
due to the nonlinearity of the semiclassical part of the Hamiltonian constraint, and

\begin{equation}
\label{ERROR22}
\xi_2=-{\Lambda \over 8}\nabla^{eb}_{af}\epsilon_{ae}\epsilon_{bf}
\end{equation}

\noindent
We can now compute the necessary terms.  
\noindent
(iii) Rewrite the quadratic contribution to $\xi_{1}$ in the following form

\begin{eqnarray}
\label{VTENSOR}
I^{ae}_{bf}\epsilon_{ae}\epsilon_{bf}=
I^{ae}_{bf}\hat{U}_{ae}^{a^{\prime}e^{\prime}}\hat{U}_{bf}^{b^{\prime}f^{\prime}}\zeta_{a^{\prime}e^{\prime}}\zeta_{b^{\prime}f^{\prime}}
\equiv\hat{V}^{a^{\prime}b^{\prime}}_{e^{\prime}f^{\prime}}\zeta_{a^{\prime}e^{\prime}}\zeta_{b^{\prime}f^{\prime}}.
\end{eqnarray}

\noindent
This has the interpretation of a linear transformation of the deformed SU(2)-invariant single contracted epsilon tensor.  Of course, it is not invariant under the transformation 
$U$, since $U$ is more general than $SU(2)$, and also due to the matter field contribution.  We have made the identification

\begin{equation}
\label{VTENSOR1}
\hat{V}^{a^{\prime}b^{\prime}}_{e^{\prime}f^{\prime}}
={1 \over 6}\bigl(\delta^a_{e}\delta^b_{f}-\delta^a_{f}\delta^b_{e}
-{G \over 4}\delta_{ab}\widetilde{\tau}_{ef}\bigr)
U_{ae}^{a^{\prime}e^{\prime}}U_{bf}^{b^{\prime}f^{\prime}}
\end{equation}

\noindent
(iv) Rewrite the cubic contribution to $\xi_{1}$ in the following form

\begin{eqnarray}
\label{ETENSOR}
\hbox{det}\epsilon={1 \over 6}\epsilon_{abc}\epsilon_{def}
\epsilon_{ad}\epsilon_{be}\epsilon_{cf}\nonumber\\
=\epsilon_{abc}\epsilon_{def}\hat{U}_{ad}^{a^{\prime}d^{\prime}}\hat{U}_{be}^{b^{\prime}e^{\prime}}\hat{U}_{cf}^{c^{\prime}f^{\prime}}
\zeta_{a^{\prime}d^{\prime}}\zeta_{b^{\prime}e^{\prime}}\zeta_{c^{\prime}f^{\prime}}
=\hat{E}^{a^{\prime}b^{\prime}c^{\prime}d^{\prime}e^{\prime}f^{\prime}}\zeta_{a^{\prime}d^{\prime}}\zeta_{b^{\prime}e^{\prime}}\zeta_{c^{\prime}f^{\prime}}
\end{eqnarray}

\noindent
Again, (\ref{ETENSOR}) can be viewed as a linear transformation of the double epsilon tensors out of $SU(2)_{-}$.  Taking into account the covariant derivative operator, we can 
write (in full-blown form, for completeness)

\begin{eqnarray}
\label{FULLBLOWN}
\zeta_{ab}=GQ^{\prime}_{ab}
+\delta_{a1}\delta_{b1}\bigl(G\Lambda\vert{B}\vert^{-1}\Omega_{0}[\zeta]
-(1/6)\Lambda{V}^{ab}_{ef}\zeta_{ae}\zeta_{bf}-{{\Lambda^2} \over {72}}
E^{abc}_{def}\zeta_{ad}\zeta_{be}\zeta_{cf}\bigr)\nonumber\\
-\delta_{a2}\delta_{b2}\bigl((1/2)G\Omega_{1}[\zeta]+{\Lambda \over 8}\bigl(\nabla^{eb}_{af}+\Gamma^{ab}_{ef}\bigr)\zeta_{ae}\zeta_{bf}\bigr)
\end{eqnarray}

\noindent
where we have defined

\begin{equation}
\label{DEFINITION}
\Gamma^{ab}_{ef}=U^{ab}_{a^{\prime}b^{\prime}}U^{ef}_{e^{\prime}f^{\prime}}C^{a^{\prime}b^{\prime}}_{e^{\prime}f^{\prime}}
+\nabla^{a^{\prime}b^{\prime}}_{e^{\prime}f^{\prime}}\bigl(U^{ab}_{a^{\prime}b^{\prime}}U^{ef}_{e^{\prime}f^{\prime}}\bigr)
\end{equation}

\noindent
The sophisticated form of the constraint has a convenient interpretation.  Note that (\ref{FULLBLOWN}) is exact to all orders and no approximation has yet been made.  Think of $\zeta_{ab}$ as a vector in an abstract nine-dimensional space that corresponds to the full solution of quantum gravity.  The right hand side can be thought of as a transformation of this vector to a new point in this space.  It is considerably more convenient to perform an iterative solution in this way, since it is easy to keep track of the $n^{th}$ iteration by computer for arbitrarily high $n$.  One simply makes note of the orbit of this vector under the generalized linear transformation and tests for various properties and measures of convergence and stability.\par
\indent
Let us assume that $\zeta_{ab}$ has been found to the order desired.  Then one can construct the CDJ matrix from this.  It is given, simply by

\begin{equation}
\Psi_{ab}=-{6 \over \Lambda}\delta_{ab}+U_{ab}^{cd}\zeta_{cd}
\end{equation} 

\noindent
Since the CDJ matrix can be thought of as an inverse cosmological constant (field-valued `tensor'), the process of iteration of (\ref{FULLBLOWN}) can be viewed as a renormalization of this `generalized' cosmological constant: a renormalization due to quantum gravitational effects, which manifest themselves at the classical level.
\par
The general solution can then be written as a power series in the cosmological constant, in the form

\begin{equation}
\label{SOLUT}
\zeta_{ab}={6 \over \Lambda}\sum_{n=1}^{\infty}(G\Lambda)^{n}\hat{U}_{ab}^{a_{1}b_{1}a_{2}b_{2}...a_{n}b_{n}}\prod_{k=1}^{n}Q^{\prime}_{a_{k}b_{k}}
\end{equation}

\noindent
$\zeta_{ab}$ in (\ref{SOLUT}) has the interpretation of a renormalization of the source 
$Q^{\prime}_{ab}$ in an infinite series in powers of a dimensionless coupling constant
$G\Lambda$.  The multi-indexed operator $\hat{U}_{ab}^{A}$ consists of products of $\hat{U}$ which translate the respective sources acted upon within the generalized manifold comprising the base space.  Each such effect is suppressed by a factor of $G\Lambda$.  The operator also contains powers of the spatial part of the matter energy momentum tensor $T_{ij}$, which is separate from the components contained in the source $T_{00}$ and $T_{0i}$.  By the use of the Einsteins equations

\begin{equation}
T_{\mu\nu}\propto{G}_{\mu\nu}
\end{equation}

\noindent
the power series (\ref{SOLUT}) can be expressed as a power series of positive powers of the Ricci curvature tensor $R_{\mu\nu}$ and the curvature scalar $R=g^{\mu\nu}R_{\mu\nu}$ and negative powers of the Ashtekar curvature $\hbox{det}B$.  The full CDJ matrix can then be written in the form

\begin{equation}
\label{SOLUT1}
\Psi_{ab}=-{6 \over \Lambda}\Bigl(\delta_{ab}+\sum_{n=1}^{\infty}(G\Lambda)^{n}
\hat{U}_{ab}^{a_{1}b_{1}a_{2}b_{2}...a_{n}b_{n}}\prod_{k=1}^{n}Q^{\prime}_{a_{k}b_{k}}\Bigr)
{\sim}-{6 \over \Lambda}\Bigl[e^{G\Lambda\hat{u}Q^{\prime}}\Bigr]_{ab}.
\end{equation}

\noindent
The compact notation on the right hand side of (\ref{SOLUT1}) signifies a new kind of generating function or group action for the asymptotic expansion, in which the `generator' $\hat{u}$ acts on the 
`parameter' $Q^{\prime}$ in a kind of tensor representation.  The specific form of the $n^{th}$ iteration of this generator can be explicitly carried out based on the entities specifying the matter model in queston.  One can then make the identification

\begin{eqnarray}
\label{EXPANSION}
{1 \over {n!}}\prod^{n}\hat{u}
=\hat{U}_{ab}^{a_{1}b_{1}a_{2}b_{2}...a_{n}b_{n}}.
\end{eqnarray}

\noindent
In this way $\hat{u}$ generates all of the propagators and vertices of the expansion.

\section{Wavefunction of the universe: Discussion}

\subsection{Mixed partials condition as a boundary condition on the wavefunction of the universe}

\noindent
The mixed partials condition is a consistency condition on the quantization procedure which must be consistent with the solution to the quantum constraints and can be viewed as arising from the commutator of the conjugate momenta, as in  

\begin{eqnarray}
\label{CONSTRAIN4}
\bigl[\hat{\widetilde{\sigma}^i_a}(x),\hat{\pi}_{\alpha}(x)\bigr]=0~\forall{x}.
\end{eqnarray}

\noindent
Equation (\ref{CONSTRAIN4}) is simply the condition that the gravitational and the matter momenta, independent dynamical variables, have trivial commutation relations with one another.  This implies, in the Schr\"odinger representation, that

\begin{eqnarray}
\label{CONSISTENT}
\hbar{G}{\delta \over {\delta{A^a_i(x)}}}
(-i\hbar){\delta \over {\delta\phi^{\alpha}(x)}}\Psi_{GKod}
=(-i\hbar){\delta \over {\delta\phi^{\alpha}(x)}}
\hbar{G}{\delta \over {\delta{A^a_i(x)}}}\Psi_{GKod}.
\end{eqnarray}

\noindent
Plugging in the CDJ Ansatz into (\ref{CONSISTENT}) results in the condition

\begin{eqnarray}
\label{CONSTRAIN5}
\Bigl[\Psi_{ae}B^i_e{\pi}_{\alpha}-{\pi}_{\alpha}\Psi_{ae}B^i_e
+\delta^{(3)}(0)\Bigl(\hbar{G}{{\partial\pi_{\alpha}} \over {\partial{A^a_i}}}
+i\hbar{B}^i_{e}{{\partial\Psi_{ae}} \over {\partial\phi^{\alpha}}}\Bigr)\Bigr]\Psi_{GKod}=0
\end{eqnarray}

\noindent
The semiclassical part of (\ref{CONSTRAIN5}) does not lead to anything new, however the vanishing of the singular term implies the condition that

\begin{eqnarray}
\label{CONSTRAIN6}
{{\partial\pi_{\alpha}} \over {\partial{A^a_i}}}
=-{i \over G}B^i_{e}{{\partial\Psi_{ae}} \over {\partial\phi^{\alpha}}}
={i \over G}B^i_{e}{{\partial\epsilon_{ae}} \over {\partial\phi^{\alpha}}}
\end{eqnarray}

\noindent
whereupon the semiclassical matter momentum $\pi_{\alpha}$ is directly related to the CDJ deviation matrix $\epsilon_{ae}$.  One can now integrate each $\alpha$ component of (\ref{CONSTRAIN6}) in the functional space of fields with respect to $A^a_i$ to obtain more conveniently in CDJ variables\footnote{This makes use of the nonconvential calculus in \cite{EYO} in which spatial and functional integration and differentiation are commuting processes.}

\begin{eqnarray}
\label{CONSTRAIN7}
\pi_{\alpha}=f_{\alpha}[\vec{\phi}]+{i \over G}
\Bigl(\int_{\Gamma}\delta{A}^a_{i}B^i_{e}\Bigr){\partial \over {\partial\phi^{\alpha}}}\epsilon_{ae}
=f_{\alpha}[\vec{\phi}]+{i \over G}
\Bigl(\int_{\Gamma}\delta{X}^{ae}\Bigr){\partial \over {\partial\phi^{\alpha}}}\epsilon_{ae}[A^b_j,\phi^{\alpha}].
\end{eqnarray}

\indent
In (\ref{CONSTRAIN7}) $f_{\alpha}[\vec{\phi}]$ plays the role of a constant of functional integration with respect to $A^a_i=A^a_{i}(x)$ and depends entirely upon the matter fields.  There are no a-priori restrictions on these functions, but it is convenient to prescribe it as a sort of boundary condition on quantum gravity with a physical interpretation.  The function $f_{\alpha}$ can be thought of as the conjugate momentum for the matter fields in the absence of gravity.  One could thus choose $f_{\alpha}$ to correspond to the conjugate momentum for the matter field quantized in the absence of gravity ($\hbox{lim}_{G\rightarrow{0}}$).  If one takes the governing quantum theory to be quantum field theory in flat Minkowski spacetime, then one can impose the observable semiclassical limit for this theory as a boundary condition on the gravity-coupled quantum theory.\footnote{By the same token, the self-duality condition can be see as a boundary condition on the coupled theory in the absence of matter fields.}  The solution to the constraints then extrapolates this boundary condition nonlinearly as a of backreaction on DeSitter spacetime.\par
\indent
There are two commuting processes occuring in (\ref{CONSTRAIN7}).  There is partial differentiation with respect to the functional dependence of $\epsilon_{ae}$ in the direction of the matter variables $\phi^{\alpha}$, as well as integration in the direction of the gravitational variables $A^a_i$.\par

\subsection{Input from the semiclassical limit into $\Psi_{GKod}$}

\noindent
The canonically determined wavefunction for the generalized Kodama state can be written, in the reduced phase space quantization scheme, as the exponential of the starting Lagrangian evaluated on the solution to the constraints \cite{EYO}.

\begin{eqnarray}
\label{WAVEFUNCTION}
\Psi_{GKod}[X^{ae},\phi^{\alpha}]=\hbox{exp}\int^{T}_{t_0}{dt}\Bigl[(\hbar{G})^{-1}\int_{\Sigma}d^3{x}\Psi_{ae}B^i_e\dot{A}^a_i
+{i \over \hbar}\int_{\Sigma}d^3{x}\pi_{\alpha}\dot{\phi}^{\alpha}\Bigr].
\end{eqnarray}

\noindent
When the quantum constraints are solved, the semiclassical-quantum correspondence is identically satisfied and the wavefunction corresponding to reduced phase space and Dirac quantization must be the same\footnote{This is a conclusion drawn by the current author, based on the results of \cite{EYO} and \cite{TQFT4}.}.  It is clear, in the solution of the constraints for a matter-coupled theory, that both the CDJ matrix $\Psi_{ae}$ and the semiclassical matter conjugate momentum $\pi_{\alpha}$ must contain dependence both upon the gravitational and the matter configuration space variables.  This is a natural consequence of the interaction between gravity and matter.\par
\indent  
In order to acquire some intuition on the functional boundary conditions on the wavefunction of the universe it is instructive to deduce the form of the wavefunction 
(\ref{WAVEFUNCTION}) in the limit when there is no interaction.  There are two extremes: First for pure gravity with cosmological term and no matter fields present the wavefunction should reduce to the pure Kodama state.  This can be most directly seen in CDJ variables from the solution of the corresponding Schr\"odinger equation

\begin{eqnarray}
\label{PUREKOD}
\Bigl[{\Lambda \over 6}\hbar^3{G}^3\epsilon_{ijk}\epsilon^{abc}
{\delta \over {\delta{A}^a_i(x)}}{\delta \over {\delta{A}^b_j(x)}}{\delta \over {\delta{A}^c_k(x)}}
+\hbar^2{G}^2\epsilon_{ijk}\epsilon^{abc}{\delta \over {\delta{A}^a_i(x)}}{\delta \over {\delta{A}^b_j(x)}}B^k_c(x)\Bigr]\Psi_{GKod}=0
\end{eqnarray}

\noindent
which leads to a nontrivial solution for the chosen operator ordering of momenta to the left of the coordinates

\begin{eqnarray}
\label{PUREGRAV}
\Psi_{GKod}[A,\vec\phi]=\Psi_{Kod}[A]=e^{-6(\hbar{G}\Lambda)^{-1}I_{CS}}.
\end{eqnarray}

\noindent
The pure Kodama state has semiclassical orbits that correspond to DeSitter spacetime \cite{KOD},\cite{POSLAMB}.\par
\indent
The other extreme is the case when there is pure matter in the absence of gravity.  Then one might expect the wavefunction of the universe to reduce to the ground state of the quantized matter theory in the Schr\"odinger representation in Minkowski spacetime.  In this case, the function 
$\pi_{\alpha}[A,\vec{\phi}]\rightarrow{f}_{\alpha}(\vec\phi)$ serves as the conjugate momentum corresponding to the WKB state.  Take for simplicity the massive Klein--Gordon field in Minkowski spacetime.  The time evolution of the state is governed in the Schr\"odinger picture, by the functional Schr\"odinger equation

\begin{equation}
\label{KLEIN}
i\hbar{\partial \over {\partial{t}}}\Psi_{KG}[\phi]=H[\phi,-i\delta/\delta\phi]\Psi_{KG}[\phi]=E\Psi_{KG}[\phi].
\end{equation}

\noindent
where the Hamiltonian operator is given by

\begin{equation}
\label{KLEIN1}
\hat{H}={1 \over 2}\int{d^3{x}}\Bigl[-\hbar^2{\delta \over {\delta{\phi(x)}}}{\delta \over {\delta{\phi(x)}}}-\phi(x)\bigl({\nabla}^2+m^2\bigr)\phi(x)
+V(\phi(x))\Bigr].
\end{equation}

\noindent
where $E$ is the energy eigenvalue of the state.  Let us set the self-interaction potential $V(\phi)$ to zero for simplicity.  Hamiltonians quadratic in the dynamical variables allow for a simple expression for the quantum mechanical wavefunction $\Psi[\phi]$ for the state in terms of a mode number basis of harmonic oscillators.  As is known, the ground state in this basis corresponds to a Gaussian.  The corresponding ground state for $\Psi[\phi]$ is given by

\begin{equation}
\label{KLEIN3}
\Psi_{KG}[\phi]=\hbox{exp}\Bigl[-{1 \over 2}\int{d^3{x}d^3{y}}\phi(x)G(x,y)\phi(y)\Bigr]
\end{equation}

\noindent
where the kernel $G(x,y)$ satisfies the relations

\begin{equation}
\label{KLEIN4}
G^2(x,y)=\bigl(-{\nabla}^2+m^2\bigr)\delta(x-y);~~G^{-1}(x,y)
=\int{d^3{p} \over {2\pi}^3}{{e^{ip.(x-y)}} \over {\sqrt{{\vert{\vec p}\vert}^2+m^2}}}.
\end{equation}

\noindent
The energy eigenvalue $E$ is just the trace of the kernel $G(x,y)$, given by

\begin{equation}
\label{KLEIN6}
E={1 \over 2}\int{d^3{x}}\int{d^3{p} \over {2\pi}^3}\sqrt{{\vert{\vec p}\vert}^2+m^2},
\end{equation}

\noindent
the zero-point energy of the scalar field $\phi(x)$.  The function $f$ in this case is determined by acting on the state 

\begin{eqnarray}
\label{KLEIN7}
\hat{\pi}(x)\Psi_{KG}=-i\hbar{\delta \over {\delta\phi(x)}}\Psi_{KG}=\Bigl[\int_{\Sigma}d^3{y}G(x,y)\phi(y)\Bigr]\Psi_{KG}=f[\phi(x)]\Psi_{KG}.
\end{eqnarray}

\noindent
with a more complicated functional form for $f$ when the potential $V(\phi)$ is nonharmonic.  The form can be deduced from the effective action for a quantized theory in flat Minkowski spacetime, either by some kind of expansion of the solution to (\ref{KLEIN1}), or via the path integration approach

\begin{equation}
\label{PATHPOOT}
\Psi_{matter}[\phi(T)]=e^{(i/\hbar)\Gamma_{eff}[\phi(T)]}
=\int{D\phi}e^{-S[\phi]}.
\end{equation}

\noindent
For a harmonic potential $V=(1/2)m^2\phi^2$ the effective action can be computed expactly from (\ref{PATHPOOT}) to be a Gaussian, which can then be expressed in terms of the boundary values of the classical solution.  In the more general case of a nonharmonic potential one resorts to a perturbative loop expansion about a reference configuration of the field 
$\phi(x,t)$, usually chosen to correspond to a stationary point of the action $S[\phi]=S[\phi_{cl}]=S_{cl}[\phi(T)]$, where $T$ is the time corresponding to the final conditions of the classical solution $\phi_{cl}$ as defined on the final hypersurface $\Sigma_T$

\begin{eqnarray}
\label{PATHPOOT1}
Z(J)=e^{iW(J)}=\bigl(\hbox{Det}(\nabla^2-m^2)\bigr)^{-1/2}e^{-S_{cl}[\phi(x,T)]}\hbox{exp}\Bigl[\int{d^4}x~V[\delta/\delta{J}(x)]\Bigr]\nonumber\\
\hbox{exp}\Bigl[{1 \over 2}\int\int{d}^4{x}d^4{y}J(x)\Delta(x-y)J(y)\Bigr].
\end{eqnarray}

\noindent
One then computes the value of $f$ via the identification 
$\Gamma_{eff}=-W(J)+\int{J}\cdot\phi$, and consequently

\begin{eqnarray}
\label{PATHPOOT1}
-i\hbar{\delta \over {\delta\phi(x)}}\Psi_{matter}[\phi]
=-i\hbar{\delta \over {\delta\phi(x)}}e^{(i/\hbar)\Gamma_{eff}[\phi(T)]}
=\Bigl({{\delta\Gamma_{eff}} \over {\delta\phi(\boldsymbol{x},T)}}\Bigr)
\Psi_{matter}[\phi]
\end{eqnarray}

\noindent
whereupon one makes the 
identification $f(\phi)=\delta\Gamma_{eff}/\delta\phi$.  This is the semiclassical limit forming the input into the quantities $Q^{\prime}_{ae}$.  This result can be applied to a wide class of models more general than the Klein--Gordon scalar field.\par
\indent
When the Klein--Gordon field is coupled to gravity, the corresponding Schr\"odinger equation takes a more complicated form given which can be written as

\begin{eqnarray}
\label{COMBINED}
\Bigl[{1 \over 6}\bigl(\Lambda+{1 \over 2}m^2\phi^2+V(\phi)\bigr)\hbar^3{G}^3\epsilon_{ijk}\epsilon^{abc}
{\delta \over {\delta{A}^a_i(x)}}{\delta \over {\delta{A}^b_j(x)}}{\delta \over {\delta{A}^c_k(x)}}
+\hbar^2{G}^2\epsilon_{ijk}\epsilon^{abc}{\delta \over {\delta{A}^a_i(x)}}{\delta \over {\delta{A}^b_j(x)}}B^k_c(x)\nonumber\\
-{{\hbar^2} \over 2}{\delta \over {\delta\phi(x)}}{\delta \over {\delta\phi(x)}}
+{1 \over 2}\partial_{i}\phi(x)\partial_{j}\phi(x){\delta \over {\delta{A}^a_i(x)}}{\delta \over {\delta{A}^a_j(x)}}\Bigr]\Psi_{GKod}=0
\end{eqnarray}

\noindent
The imprints of the scalar field model are rearranged when quantized on the same footing as gravity, but one should hope that the solution to (\ref{COMBINED}) should reduce to that for (\ref{PUREKOD}) in the limit of pure gravity and to that for (\ref{KLEIN}) in the limit of pure matter.  This provides a physical basis for fixing the constant of integration $f_{\alpha}$ in the mixed partials condition corresponding to (\ref{COMBINED}), which is given by \cite{EYO}

\begin{eqnarray}
\label{MIXEDKG}
\pi(x)=f(\phi(x))+{i \over G}\Bigl(\int_{\Gamma}\delta{X}^{ae}(x)\Bigr){\partial \over {\partial\phi(x)}}\epsilon_{ae}[X^{bf}(x),\phi(x)]~\forall{x}.
\end{eqnarray}

\indent
To make the boundary conditions more apparent one can expand the conjugate momentum variables in (\ref{WAVEFUNCTION}) about the corresponding boundary conditions.  This is given by

\begin{eqnarray}
\label{EXPANN}
\Psi_{ae}=-\bigl({6 \over \Lambda}\delta_{ae}+\epsilon_{ae}\bigr);
~~\pi_{\alpha}=f_{\alpha}+{i \over G}\Bigl(\int{B}^i_e\delta{A}^a_i\Bigr){\partial \over {\partial\phi^{\alpha}}}\epsilon_{ae}.
\end{eqnarray}

\noindent
As long as these expansions (\ref{EXPANN}) mutually consistent, then the resulting wavefunction should automatically incorporate the appropriate semiclassical limit into the coupled quantum theory.  The source terms in the quantum constraints arising from $f_{\alpha}$ have the interpretation of a backreaction on the spacetime determined by the pure Kodama state, which is DeSitter spacetime.  Substituting (\ref{EXPANN}) into 
(\ref{WAVEFUNCTION}), one obtains

\begin{eqnarray}
\label{WAVEFUNCTION1}
\Psi_{GKod}[X,\phi]=e^{-6(\hbar{G}\Lambda)^{-1}I_{CS}}e^{{i \over \hbar}\int_{\Sigma}\vec{f}\cdot\delta\vec{\phi}}
\hbox{exp}\int^T_{t_0}\Bigl[-2(\hbar{G})^{-1}\int_{\Sigma}d^3{x}~\delta{X}^{ae}\epsilon_{ae}[X(x),\phi^{\alpha}(x)]_f\Bigr]
\end{eqnarray}

\noindent
Equation (\ref{WAVEFUNCTION1}) consists of a pure gravitational ground state $\Psi_{GKod}$, a matter part which drives statistical quantum fluctuations from this ground state, which is given by the wavefunction for the quantized matter field in the absence of gravity and a part encoding the backreaction effects as determined by the solution to the constraints, which is labeled by the semiclassical matter momentum $\vec{f}=f_{\alpha}(\phi)$.  The function forms an input via the inhomogeneous source vector in the constraint equations

\begin{eqnarray}
\label{SOURCES}
H_i=f_{\alpha}(D_i\phi)^{\alpha};~~Q_a=f_{\alpha}(T_a)^{\alpha}_{\beta}\phi^{\beta};
~~\Omega_0=\widetilde{\tau}_{00}(f);~~\Omega_1=-{i \over {4G}}\sum_{\alpha}{{\partial{f}_{\alpha}} \over {\partial\phi^{\alpha}}};~~\Omega_2=\Omega_2(f).
\end{eqnarray}

\subsection{Generalized Kodama state for the full theory}

\noindent
The generalized Kodama state, modulo prefactors due to path integration, can be expressed as \cite{EYO}

\begin{eqnarray}
\label{FULWAV}
\Psi_{GKod}[A,\phi]=\hbox{exp}\Bigl[\int_{M}\bigl((\hbar{G})^{-1}\Psi_{ae}B^i_{e}\dot{A}^a_i
+{i \over \hbar}\pi\dot\phi\bigr)\Bigr]\biggl\vert_{C_{ab}=0}
\end{eqnarray}

\noindent
Incorporating the mixed partials condition into the matter momentum \cite{EYO},

\begin{eqnarray}
\label{MIX}
\pi(x)=f[\phi(x)]-{i \over G}{\delta \over {\delta\phi}}\int_{\Gamma}\delta{X}^a_e\Psi_{ae}(x)
\end{eqnarray}

\noindent
We have switched from $\partial$ to $\delta$ notation regarding the matter field for consistency with the functional one-form $\delta\phi$.  The reader should bear in mind that all field-theoretical infinities have already been factored out, so $\delta/\delta\phi(x)$ is really a partial derivative with respect to $\phi$ at the point $x$ and not a functional derivative.

\begin{eqnarray}
\label{FULLWAVE}
\Psi_{GKod_f}[A,\phi]=e^{-6(\hbar{G}\Lambda)^{-1}I_{CS}[A]}
\hbox{exp}\Bigl[{i \over \hbar}\int_{\Sigma}d^3{x}\int_{\Sigma}f[\phi]\delta\phi\Bigr]\nonumber\\
\hbox{exp}\Bigl[(\hbar{G})^{-1}\int_{\Sigma}d^3{x}\int_{\Gamma}\delta{X}^a_e\Bigl[1+\delta\phi
{\delta \over {\delta\phi}}\Bigr]\Psi_{ae}[X^b_f,f,\phi]\Bigr]
\end{eqnarray}

\noindent
So we see that there is a contribution to the gravitational sector from the matter field due to the mixed partials condition.  To see more clearly the nature of this contribution, note that there are two operations occuring at each spatial point $x$.  There is differentiation with respect to $\phi$, holding $A^a_i$ fixed, followed by integration with respect to $\phi$, again at fixed $A^a_i$.  This looks like a minisuperspace effect, but is in fact still the full theory.  The differentiation and integration over the functional space of $\phi$ are inverse operations to each other, and thus cancel each other out.  The next result is that the matter momentum makes a contribution to the wavefunction exactly matching that from the gravitational sector.  Equation (\ref{FULWAV}) leads, 
applying \ref{SOLUT1} to the results of \cite{EYO}, to

\begin{eqnarray}
\label{FULLWAVE1}
\Psi_{GKod_f}[A,\phi]=\Psi_{Kod}[A]\hbox{exp}\Bigl[{i \over \hbar}\int_{\Sigma}d^3{x}\int_{\Sigma}f[\phi]\delta\phi\Bigr]\nonumber\\
\hbox{exp}\Bigl[(12\hbar{G}\Lambda)^{-1}\int_{\Sigma}d^3{x}\int_{\Gamma}
\delta{X}^{ab}\sum_{n=1}^{\infty}(G\Lambda)^{n}
\hat{U}_{ab}^{a_{1}b_{1}a_{2}b_{2}...a_{n}b_{n}}
\prod_{k=1}^{n}Q^{\prime}_{a_{k}b_{k}}[A,\phi,f]\Bigr]
\end{eqnarray}

\noindent
There are a few things of note regarding the generalized Kodama state: (i) It is labeled by $f$, a function purely of the matter fields, as a nonseparable basis \cite{EYO}.  The function $f$ can be thought of as the semiclassical matter momentum in the absence of gravity, and can be chosen such as to impose the proper semiclassical limit of the fully coupled theory; (ii) The gravitational sector is expressible as an asymptotic expansion, on the final spatial hypersurface $\Sigma_T$, relative to the pure Kodama state in powers of $G\Lambda$; (iii) the functional integration over the space of fields $\Gamma$ occurs at each fixed position $x$ on $\Sigma_T$ to produce a functional of the fields which is then integrated over position $d^3x$; (iv) The operators 
$\hat{U}_{ab}^{a_{1}b_{1}a_{2}b_{2}...a_{n}b_{n}}$ act on the functional dependence of the matter source terms $Q^{\prime}_{ab}\sim(T_{0i},Q_a,q)$ upon the Ashtekar connection $A^a_i$; (v) The asymptotic expansion is in terms of products of the local matter matter $SU(2)\otimes{Diffeo}\otimes{H}$ charges propagated around $B_{Kod}$, with coefficients involving $T_{ij}$.  This might be preferable, due to the dimensionless coupling constant, to an expansion in powers of curvature found in many higher-derivative theories of gravity.

\subsection{Network analogy to Feynman diagrams}

There is another interpretation available for the series expansion (\ref{FULLWAVE1}).  If one thinks of the source $Q^{\prime}_{ab}=Q^{\prime}_{ab}[x,A,\phi]\sim{q}^{\prime}$ as a 
matter-dependent property smeared over the base space $B_{Kod}$ of $E_{Kod}$ and the operator $\hat{U}$ as a generalized propagator and the error as a self-interaction vertex 
parametrized by the coupling constant $(G\Lambda)^n$, then one can visualize the following process in which the source traces out an orbit which can be projected onto the base 
space $B_{Kod}$.\par
\indent  
The source $q^{\prime}=q^{\prime}_{ab}$ propagates a disturbance in functional space from $\chi_0$ to $\chi_1$ where it undergoes a scattering at a 3-point and 4-point vertex.  The resulting disturbance combines with the source already present at $\chi_1$ to form a new source $q^{\prime\prime}$.  This new source propagates from $\xi_1$ to $\xi_2$ where it undergoes another scattering  and combines with the source present at $\xi_2$ to form the source $q^{\prime\prime\prime}$ and the process repeats itself.  The number highest of source particles at a particular stage proliferates as $3^n$ and the total matter source, $q^{\prime\prime...\prime}$ behaves as though it is undergoing a kind of random walk through $B_{GKod}$.  The walk is not really random, it is predetermined by the dynamics of the quantized constraints.  At each stage the source accumulates more and more particles.  One hopes that the total source to infinite order should be convergent.

\end{document}